# Critical Structural Parameter Determining Magnetic Phases in the $Fe_2Mo_3O_8$ Altermagnet System

T. A. Tyson[1,3,*], S. Liu[1], S. Amarasinghe[1], K. Wang[2,3],
S. Chariton[4], V. Prakapenka[4], T. Chang[4], Y.-S. Chen[4], Z. Liu[5]
C. J. Pollock[6], S.-W. Cheong[2,3], and M. Abeykoon[7,*]

[1]Department of Physics, New Jersey Institute of Technology, Newark, NJ 07102, USA
[2]Department of Physics and Astronomy, Rutgers University, Piscataway, NJ 08854, USA
[3]Rutgers Center for Emergent Materials, Rutgers University, Piscataway, NJ 08854, USA
[4]Center for Advanced Radiation Sources, The University of Chicago, Argonne, IL 60439, USA
[5]Department of Physics, University of Illinois at Chicago, Illinois 60607-7059, USA
[6]Cornell High Energy Synchrotron Source, Wilson Laboratory, Cornell University, Ithaca, New York 14853, USA
[7]National Synchrotron Light Source II, Brookhaven National Laboratory, Upton, NY 11973, USA

*Corresponding Authors:

T. A Tyson, e-mail: tyson@njit.edu

M. Abeykoon, e-mail: aabeykoon@bnl.gov

# Abstract

A systematic structural study of the $Fe_2Mo_3O_8$ system as a function of pressure, temperature, and magnetic field reveals that the $P6_3mc$ space group of this material remains stable for a broad range of these parameters. No changes are seen in the long-range structure for pressures between 0 and 10 GPa, temperatures between 11 K and 300 K, and magnetic fields up to 9 T. The magnetostructural response ($\Delta c/c$) for a magnetic field transverse to the c-axis displacement is determined. The system is found to exhibit strong magnetostructural coupling. The well-known magnetic-field-induced first-order transition is found to be isostructural, between two distinct altermagnet states, and is accessible by fields in the a-b plane. In terms of the c/a ratio, the structures between ambient pressure and 10 GPa are found to map onto the full Zn doping range ($(Fe_{1-y}Zn_y)_2Mo_3O_8$, $0 \leq y \leq 1$) in this system. The results show that the critical sensitive structural parameter for tuning the magnetic properties with pressure, temperature, and pressure is the c-axis length. The results indicate that the magnetic order in this complex metal oxide system (A2Mo3O8, A=Co, Mn, and Ni) can be cleanly tuned by pressure, making this class of materials an excellent platform for magnetic order switching in films grown on piezoelectric substrates.

# I. Introduction

The examination of the coupling of spin, lattice, and electron degrees of freedom in complex correlated oxide systems (multiferroics) has been an active area of research for several decades [1,2,3,4,5,6,7,8]. The characteristic features that make these materials relevant for device application include low power storage and high sensitivity, enabling small domain size and high-density storage. The possibility of beyond-binary (0/1) information storage is also envisioned. Combining the properties of antiferromagnetic and ferromagnetic systems, altermagnet systems characterized by broken PT (parity times time reversal) symmetry [9], having zero net magnetization (full compensation) but hosting magnetic atoms in distinct chemical environments promise to enable new magnetic sensing and storage capabilities [10,11,12,13,14]. $Fe_2Mo_3O_8$ is studied to understand the structural tuning of magnetic properties.

The polar magnet $Fe_2Mo_3O_8$ came under extensive study earlier due to the observation of large magnetoelectric coefficients (quantifying the coupling of magnetization and electric polarization) [15,16]. It is currently the single-phase material with the largest magnetically switchable electric polarization. Specifically, this system exhibits a change in electric polarization of $\approx$ 0.3 μC/cm$^2$ on cooling below the Néel temperature ($\approx$ 60 K). Below the antiferromagnetic ordering temperature, the application of magnetic fields (along the c-axis) induces jumps in the sample magnetization concomitant with abrupt changes in the sample electric polarization ($\approx$ 0.1 μC/cm$^2$ jumps). At 55 K, a step change in magnetization (increase in magnitude) is found between 4.5 and 6 T. This step change is related to an antiferromagnetic (AF) to ferrimagnetic (FRM) transition in which the distinct spins on pairs of tetrahedral $FeO_4$ sites (Fe1 sites) and pairs of octahedral $FeO_6$ sites (Fe2 sites) switch from being antiparallel to being parallel with resulting in a residual net magnetization. This transition produces a jump in polarization of approximately 0.13 μC/cm$^2$. While magnetization loops exhibit extremely weak hysteresis, polarization loops show clear hysteresis. Near 3 T at 55 K (AF region), a linear magnetic magnetization response to external electric fields is observed. The magnetic response to electric fields is an order of magnitude larger than that found in the polar magnet $Ni_3TeO_6$. Doping $Fe_2Mo_3O_8$ with Zn stabilizes the FRM phase at the expense of the AF phase [10,17,18,19,20].c For $y \leq 1/2$ (in $(Fe_{1-y}Zn_y)_2Mo_3O_8$), the Zn ions preferentially occupy the tetrahedral sites (Fe1 replacement) [21,22]. Mössbauer measurements on the Zn-doped (y=0.5) and the undoped systems (y=0) reveal a 2+ valence on all metal sites (Fe or Zn) and also indicate preferential occupancy of the tetrahedral sites by Zn [23]. A lambda-shaped feature occurs in the heat capacity, which is typical of magnetic ordering [15] transitions. The heat capacity is well modeled by a simple Debye model at high temperatures, but this behavior is not recovered at low temperatures [15]. The lambda-shaped feature is found to broaden and shift to higher temperatures with the applied magnetic field [24]. The thermal Hall conductivity is $10^2$ times larger than in any system previously measured, indicating that extremely strong spin-phonon coupling exists in this material [25]. Magnetic field-induced splitting of the phonon near 50 $cm^{-1}$ and the occurrence of magnetic moments carried by the phonon are observed at low temperatures [26]. We note that for all of these measurements, the magnetic field is applied along the c-axis. We will show that changes for magnetic fields applied in the a-b plane can recover structural responses compatible with the c-axis field-dependent measurements.

Density functional theory simulations suggest a structural transition from the paramagnetic to the antiferromagnetic phase and from the field-induced antiferromagnetic to the ferrimagnetic phase [27]. These theoretical calculations suggest that oxygen displacements induced by magnetic ordering determine the electric polarization that lies along the c-axis. Understanding the changes in structure with temperature, pressure, and magnetic field is important to optimize this system for practical use.

Under ambient conditions, the $A_2Mo_3O_8$ systems have been found to have polar space group $P6_3mc$ [28,29,30,31]. Interestingly, no high-resolution structural measurements, such as single crystal diffraction above and below the magnetic ordering temperatures, have been conducted. In this work, we show the direct structural response of this material to pressure, temperature, and magnetic field in order to assess the nature of the coupling of the lattice to the magnetic field and determine the critical structural parameters that stabilize the magnetic phases. Extracting the magnetostructural coupling will enable the development of accurate theoretical models of materials in this class and point to the prediction of systems with large coupling with low field response, possibly near room temperature. The critical structural parameter that drives the magnetic order is found to be the c lattice length. Temperature-dependent single-crystal X-ray diffraction measurements above and below the magnetic transition were conducted.

No changes are seen in the long-range structure for pressures between 0 and 10 GPa, temperatures between 11 K and 300 K, and magnetic fields up to 9 T. The previously observed field-induced first-order transition is found to be isostructural and is accessible by fields in the a-b plane. This transition corresponds to switching between two distinct magnetic phases [9]. In terms of the c/a ratio, the structures between ambient pressure and 10 GPa are found to map onto the full Zn doping range ($(Fe_{1-y}Zn_y)_2Mo_3O_8$, $0 \leq y \leq 1$) in this system. The system is found to exhibit strong magnetostructural coupling, which is expected in altermagnet materials.

# II. Experimental Methods

Crystals of $Fe_2Mo_3O_8$ were prepared by chemical vapor transport [15]. More complete details on the experimental methods, modeling, and analysis are provided in the supplementary document. Diamond cell preparation, testing and sample loading were conducted at NSLS2 beamline 22-IR-1. Temperature-dependent single crystal diffraction measurements were conducted on an 18mm × 10 mm × 10 mm single crystal (at 70 K and below) at the Advanced Photon Source (APS) beamline 15-ID-D at Argonne National Laboratory using a wavelength of 0.41328 Å (30 keV High-pressure powder XRD measurements at room temperature were performed at APS beamline 13-BM-D (GESCARS) at Argonne National Laboratory using a wavelength of 0.3344 Å (37 keV). Temperature-dependent and magnetic field-dependent X-ray absorption measurements were conducted at the CHESS PIPOXS beamline ID2A at Cornell University. Temperature and magnetic field-dependent powder diffraction data were collected using Beamline 28-ID-1 (PDF) at NSLS2 (National Synchrotron Light Source) at Brookhaven National Laboratory. To determine force constants and phonon density of states for $Fe_2Mo_3O_8$, density functional calculations in the projector augmented wave approach were carried out utilizing the VASP code [32].

# III. Results and Discussion

## III. A. Single Crystal Diffraction Results

The crystal structure of $Fe_2Mo_3O_8$ was measured at room temperature and between 11 K and 70 K using synchrotron-based single-crystal X-ray diffraction methods. Highly redundant data sets were collected over the full sphere in reciprocal space with no assumption of space group symmetry. Systematic analysis of the data indicates that the same space group, $P6_3mc$, is maintained for the full temperature range. (Detailed structural information is given in Tables S1 (11 K), S2 (70 K), and S3 (300 K) of the Supplementary Information document [33]). Comparing the structure at 11 K and 70 K and utilizing an ionic model for the charge on each atom, we obtain a change in electric polarization of $\Delta P \approx 0.16$ μC/cm$^2$ between these temperatures. While the positions of the atoms (xyz fractional coordinates) within the unit cells vary slightly with temperature, more significant changes occur in the lattice parameters (see below). In terms of atomic-level changes, the dominant atomic parameter changes with temperature are in the atomic displacement parameters, $U_{ij}$. Focusing on the diagonal terms, we see that for the O2 atom at the apex of the $Fe1O_4$ tetrahedron (see Fig 1(a)), there is a significant reduction of Upara/Uperp (= $U_{33}/(U_{11}/2 + U_{22}/2)$) on passing through the magnetic ordering temperature at 60 K. This change is recovered at low temperatures. Thermal motion of O2 atoms (Fig. 1(a)) transverse to the a-b plane is enhanced relative to the motion along the c-axis. No other atom shows this response (See Fig. S1). The results suggest that anisotropy in motion exists when comparing the $a - b$ plane vs. $c$-axis, which influences the magnetic ordering. Pressure-dependent measurements will shed light on the presence of this anisotropy.

## III. B. High-Pressure X-ray Diffraction Results.

High-pressure powder X-ray diffraction measurements were utilized to extract variation in the lattice parameters ($a$ and $c$) and their ratio ($c/a$) as a function of pressure. In Fig. 2(a), we show the pressure dependence of a limited number of peaks and provide more complete images in the supplementary document (Fig. S4). There is no evidence for a structural phase change. As seen in Fig. 2(b), the c-axis is compressed more rapidly with pressure than the a-axis. For comparison, we show the ambient pressure c/a ratio in the Zn-doped system ($(Fe_{1-y}Zn_y)_2Mo_3O_8$) in the inset of Fig. 2(b) (with data from Ref. [29]). We note that the $c/a$ ratio under compression (ambient to $\approx$10 GPa) varies over the same range as the range accessible by completely doping Zn on the Fe sites in $Fe_2Mo_3O_8$ ($0 \leq y \leq 1$). Note also in the inset that doping has no significant impact on the $a$ lattice parameter. The results suggest that the critical parameter for determining the stable magnetic phases is the $c$-axis length. As a result, we now see that pressure (hydrostatic or uniaxial) can be used to tune the magnetic ground state.

A simple Murnaghan equation of state ($P = \frac{B}{B'}[(\frac{V_0}{V})^{B'} - 1]$) is valid for volume compression $V/V_0 > 0.9$ ($V_0$ is the volume at zero pressure, B is the bulk modulus, and B' is the pressure derivative of B), which is the configuration in this work. Fits to the volume ($V = a^2 b \frac{\sqrt{3}}{2}$) with this equation of state (Fig. S4(e)) yielded a bulk modulus of B = 138.4±6.9 GPa and its pressure derivative B' =10.9±2.2. The large B' value is consistent with the compression anisotropy (as can be seen by comparison with cases for hexagonal and orthorhombic $RMnO_3$ manganite systems Ref. [34]). The rate of change of the c-axis with pressure relative to the a-axis is close to 2. This gives an estimated Poisson ratio $\approx -1$ using the lattice parameters of the system. Given the c-axis's greater compressibility relative to the a-axis, either strain (in the a-b plane) or hydrostatic pressure can be used to tune the magnetic phase by adjusting the c lattice parameter.

## III. C. Temperature-Dependent Structure

## III. C1. Local Structure vs. Temperature

To understand changes in structure with temperature, local structure measurements were performed using X-ray absorption spectroscopy (XAFS). Powder diffraction measurements were conducted to assess the long-range structure. XAFS measurements based on the Fe K-edge spectra were conducted between 300 K and 5 K. These measurements enable access to the coordination shells about the average Fe site. In Fig. 3, we show the structure function (which gives the electron scattering amplitude weighted radial distribution) of the average Fe site (Fe1 and Fe2) in Fig. 3(a). The first peak in Fig. 3(a) (at 5 K) is due to the first shell Fe-O bond, while the second peak is a combination of the higher-order Fe-O shells, the nearest and the next nearest neighbor Fe-Fe bonds, as well as the corresponding Fe-Mo shells. The surface plot (Fig. 4(b)) covering the full temperature range shows a very weak temperature dependence of the Fe-O peak but a much stronger temperature dependence of the second peak. This indicates very strong Fe1-O and Fe2-O bonds consistent with the high-frequency phonons related to these polyhedra, as shown in Fig. 6.

The radial distribution of the average Fe-O bond distance was extracted by modeling the first peak (See supplementary section for details). Example distributions are given in Fig. 4(a). The first shell Fe-O distribution exhibits clear splitting into two components at low temperatures. The surface plot of this curve for the full temperature series in Fig. 4(b) reveals that the lower-distance component of this distribution near 1.9 Å sharpens as the temperature is reduced. The sharpening becomes most profound below ~150 K. This characteristic on-set temperature (~150 K) is also seen in long-range structure measurements near a maximum in the lattice d(002) spacing due to full magnetic ordering below 60 K. It signals a precursor structure to the magnetically ordered phase significantly above the Néel temperature.

# III. C2. Long Range Structure vs. Temperature

The change in the long-range structure with temperature was determined by powder X-ray diffraction measurements. Full sphere data sets were collected at each temperature, and this enabled the extraction of lattice parameters by Rietveld refinements (See Fig. S3 and supplementary text). In Fig. 5(a), we see that below 150 K, the temperature dependence of the lattice parameters deviates from the Debye model fits [35] both for the $a$ and $c$ (black and blue curves). Also, the $c/a$ ratio (red data points) exhibits an abrupt change in slope near the magnetic ordering temperature. Changes in the average lattice parameters with temperature are similar to those previously reported in Ref. [36]. To understand the changes below 150 K, we examine the width and value of the d(002) plane separation. This is done by fitting the (002) peak in the X-ray diffraction pattern (See, for example first peak in Fig. S3(b)). The width of this peak starts to broaden near 180 K on cooling and reaches a maximum width near 80 K (blue circles in Fig. 5(b)). For lower temperatures, it sharpens continuously. This suggests structural rearrangements to support the AF phase, which initiates at high temperatures. The structural rearrangements produce two features in the length of d(002), which are shown as the red symbols. We find that the change in length (d(002)) goes through a maximum, which peaks near 140 K and then increases again on passing through the magnetic ordering transition near 60 K. The peak near 140 K is in the same region with large diffuse scattering seen by neutron scattering methods [36]. Hence, local structural rearrangement is associated with the onset of the magnetically ordered phase.

# III. D. Phonons

The phonon density of states can provide details about the motion of the atoms in the polyhedral. In Fig. 6, we see that the motion of the atoms in the $FeO_6$ octahedra (Oh) extends to the highest frequencies in the spectrum. Interestingly, the density of states for Mo (heaviest atom) has a significant amplitude at much higher frequencies than that of the Fe sites. The motion of the Fe1 and Fe2 ions covers the region between 0 and 400 $cm^{-1}$. The vibrational modes involving the octahedra and tetrahedra oxygen atoms extend to the highest frequency, with those in the octahedra sites being the most robust. This suggests that the $Fe2O_6$ polyhedra (octahedra) are more rigid than the $Fe1O_4$ tetrahedra.

## III. E. Magnetic Field Dependent Structure and Accessing Magnetic Ground States in $A_2Mo_3O_8$ Systems

## III. E1. Local Structure vs. H.

To understand the changes in the local structure with the applied magnetic field, the sample was cooled from room temperature to 55 K (in zero field), and the magnetic field was increased from 0 to 9 T and measured in 1 T steps. We establish the sensitivity of the measurements by plotting the low field 0 T and 1 T curves in blue (solid and dotted, respectively) and the high field 8 T and 9 T curves in red (solid and dotted, respectively) in Fig. 7(a). What is observed at these endpoints in the measurements is a suppression of the structure function's amplitude at high magnetic fields. Specifically, the structure-function peak amplitudes are suppressed without a change in shape in high magnetic fields. This is consistent with an increase in the mean square distribution (broadening of the radial distribution curves) with magnetic field. Note that both the Fe-O and the high-order peak are suppressed with increasing magnetic fields.

We examine the change in the magnetic field by looking at surface plots of the structure function. For the first peak (Fe-O), we see a trend in which the polyhedra remain robust up to approximately 7 T after which there is suppression (Fig. 7(b)). This suppression is due to a loss of coherence in the Fe-O interactions. A much stronger magnetic field dependence of the second peak (Fe-O (>1st shell), Fe-Fe shells, and Fe-Mo shells) is observed (Fig. 7(c)). Specifically, significant suppression is seen above 3 T. The field dependence is consistent with loss of structural coherence but without loss of symmetry. That is, the magnetic field reduces the structural domain size when passing through the magnetic field-induced transition (as expected for a first-order transition). However, the transition does not change atomic symmetry. Hence, the field-induced transition is an isostructural transition. Physically, the change in amplitude is due to a modification of the potential wells in which the atoms move.

We also examined the effect of magnetic field on the local structure from the perspective of the local structure about the Mo site. The full decomposition of the peaks (atom labels) by shell type is given in Fig. S2(b). The sensitivity of the measurements is seen by comparing the 0 T and 0.5 T curves in blue (solid and dotted, respectively) and the 7.5T and 8 T curves in red (solid and dotted, respectively) in Fig. 8(a). As in the case of the Fe-centered structural changes (Fig. 7), what is observed at the endpoints in the measurements is the suppression in the amplitude of the structure function at high magnetic fields, except for the Mo-O peak, which is associated with nonmagnetic atoms exclusively. Note that the high-order peaks beyond the Mo-O peaks are suppressed with increasing magnetic field. Again, the application of the magnetic field results in a lowering of structural coherence. (Fig S2(a) shows the raw XAFS data at 0 T and 8 T, also revealing suppression with magnetic field.)

The surface plots of the structure-function give more detail. For the second peak (Mo-Mo), we see a trend in which correlations between distinct neighboring MoOx polyhedra remain robust up to approximately 7 T after which there is a weakening of the structure function (Fig. 8(b)). A response at a much lower field (between 3 T and 4 T) is seen in the third peak (Mo-Mo and Mo-Fe Oh). The result supports changes in the tetrahedrally coordinated Fe sites (Fe2) onsetting in the region between 3 to 4 T. Overall the suppression of peak without change in the structure function shape reflects the fact that the potential wells in which the atoms move are modified by magnetic fields. This is possibly caused by charge redistribution.

# III. E2. Long-Range Structure vs. Magnetic Field

The change in the long-range structure with temperature was determined by powder X-ray diffraction measurements by examining the average results by Rietveld fits above (Fig. 5 and Fig. S3). However, directional-dependent magnetic field response can also be obtained. With the use of the hard X-ray photons (74 keV) and the incident beam parallel to the magnetic field, the special case for $c$-axis response to magnetic fields in the $a - b$ plane can be accessed by looking at low-angle peaks such as the (002) peak (Fig. 9(a) and 9(b)). We examined the change in the d(002) (see the first peak in Fig. S3(b)) for fields increasing from 0 T to 5 T and then back to 0 T. For the geometry of the experiment, only grains in the sample with the magnetic field in the a-b plane contribute to the d(002) ring on the 2D detector. Due to the high energy, the angle between the magnetic field and a-b plane is approximately 1° (θ). The change in d(002) as a function of magnetic field on increasing field and decreasing field shows strong hysteresis (Fig. 9(c)). The curves for increasing and decreasing field (covering the AF and FRM regions, separately) were fit to second-order polynomials of the form $d(H) = a1+b1*H + c1*H^2$. (Note that this experiment yields the response of the c-axis length and distribution to a-b plane magnetic fields.)

The fit for increasing fields is shown in red, and this covers the AF region leading up to the first-order transition to the FRM region. With the field increasing from zero, d(002) increases in a quadratic manner according to the function above. The d(002) spacing decreases until about 1.5 T and then increases up to a maximum value (near 3.8 T) before crossing into the FRM. In the FRM region, d(002) decreases and passes through a local maximum near 4.5 T. Following d(002) with the magnetic field being reduced from 5 T, we see a smoother quadratic-type curve, which peaks near 3.0 T. The offset in positions of the maxima from the ramping up field (3.8 T vs. 3.0 T) is characteristic of a first-order AF to FRM transition. On entering the AF region below 3.0 T (ramping down), the quadratic behavior is again followed. The c1 coefficients ($H^2$ coefficients) are similar for all regions fit. We are observing changes between the AF state and the FRM state at 55 K induced by magnetic fields in the a-b plane.

In Fig. 10, we show the change in width of the d(002) peak as the magnetic field is increased and then decreased. The large open loop (hysteresis) shows the first-order nature of the field-induced transition. The increase in width with increasing magnetic field indicated a loss of correlation due to the formation of small FRM-phase domains. This modifies the atomic potential wells, as seen in the XAFS results. We can relate this variation in d(002) to the sample's magnetostriction, which is quadratic in H.

The first-order transition occurs at the same field in which it is triggered by c-axis fields. Using the labeling conventions for altermagnetic phases by Cheong and Huang [9], we observe a change from a Type II altermagnet to a Type I altermagnet. In altermagnets (ignoring spin-orbit coupling), there is no net spin angular momentum, but there is a spin texture resulting from crystallographic variation yielding PT violation (Fe atoms with distinct structural environments such as Fe1 and Fe2). These PT variations lift Kramer degeneracy in unique parts of the Brillouin Zone away from the gamma point. Type I materials have a net magnetic magnetization (ferrimagnetic phase here) in the absence of an external perturbation. Type II materials have no net magnetization but exhibit net magnetic moments in the presence of PT symmetric perturbations (applied electric fields, thermal currents, stress). Hence, we are exploring field-induced transitions from type II to type I altermagnet phases in $Fe_2Mo_3O_8$. In the following paragraphs, we will demonstrate how the structural response is connected to the applied magnetic field.

A fourth-rank tensor $\overline{N}_{ijkl}$ defines magnetostriction connecting a material's distortion matrix (strain tensor), $\epsilon_{ij}$ , to the magnetization vector components $M_i$ via [37,38]

$$\epsilon_{ij} = \overline{N}_{ijkl} M_k M_l$$

where the sums on $k$ and $l$ are implicit. For the magnetic point groups $6'mm'$ (antiferromagnetic state (AF)) and $6m'm'$ (ferrimagnetic state (FRM)) the magnetostriction tensor $\overline{N}_{ijkl}$ takes on the same form for the point group 6mm. We then have

$$\begin{pmatrix} \epsilon_{xx} \\ \epsilon_{yy} \\ \epsilon_{zz} \\ \epsilon_{yz} \\ \epsilon_{xz} \\ \epsilon_{xy} \end{pmatrix} = \begin{pmatrix} N_{11} & N_{21} & N_{13} & 0 & 0 & 0 \\ N_{21} & N_{11} & N_{13} & 0 & 0 & 0 \\ N_{31} & N_{31} & N_{33} & 0 & 0 & 0 \\ 0 & 0 & 0 & N_{55} & 0 & 0 \\ 0 & 0 & 0 & 0 & N_{55} & 0 \\ 0 & 0 & 0 & 0 & 0 & N_{66} \end{pmatrix} \begin{pmatrix} M_x^2 \\ M_y^2 \\ M_z^2 \\ M_y M_z \\ M_x M_z \\ M_x M_y \end{pmatrix}$$

where $N_{66} = 1/2(N_{11} - N_{21})$. To apply this formalism to the $Fe_2Mo_3O_8$ system for the measurements reported here, we note that the magnetization is linear in the magnetic field for H normal to the c axis over a broad range (0 to ≈ 58 T [39]). Hence, for measurements in our experimental geometry, we have $M_x = C_1 \cos(\theta)H$, and $M_y = C_2 \sin(\varphi)H$, with H being the magnitude of the external magnetic field (and being normal to the c-axis of the crystal). We simplify the components of $\epsilon_{ij}$ and average the angular orientations $\theta$ and $\varphi$ over a full circle. For the powder sample, this gives averaged quantities: $\langle\epsilon_{xx}\rangle = H^2(C_1^2 N_{11} + C_2^2 N_{21})/2$, $\langle\epsilon_{yy}\rangle = H^2(C_1^2 N_{21} + C_2^2 N_{11})/2$, and $\langle\epsilon_{zz}\rangle = H^2 N_{31}(C_1^2 + C_2^2)/2$. For the geometry under study ($d_{002}$), we have access to the quantity $\langle\epsilon_{zz}\rangle$. The impact of this will be observed as shifts in the c-axis Bragg peaks with magnetic field. Note that $\langle\epsilon_{xy}\rangle$ gives zero when averaged over all particles. Comparing the functional forms with our data, we observe that the quadratic term in the fits in Fig. 9(c) dominates in both the AF and FRM phases, indicating that the observed response in the crystal is indeed magnetostriction. The coefficients $N_{ij}$ are expected to be unique for each structural region. The fact that they are quite similar in value (look a at c1, c1', and c parameters in Fig. (c)) here indicates that the space group is not altered by magnetic field. In fact, by examining the goodness-of-fit parameters in the field-dependent Rietveld results (Fig. S5(b)), we see that Rw is featureless as a function of magnetic field.

To address the linear terms in the fit, we consider piezomagnetism (linear magnetostriction), which is defined by a third-rank tensor $\overline{\Lambda}_{ijk}^T$ relating the materials distortion matrix (strain tensor), $\epsilon_{ij}$, to the external magnetic field vector components $H_k$ as $\epsilon_{ij} = \overline{\Lambda}_{ijk}^T H_k$. For the experimental conditions here, we have $\epsilon_{xx} = \epsilon_{yy} = \epsilon_{zz} = 0$ and $\langle\epsilon_{yz}\rangle = \langle\epsilon_{xz}\rangle = 0$. The coefficients $\Lambda_{ij}^T$ can only be accessed by single-crystal measurements. This will be addressed in future experiments.

As mentioned above, we examined the width of the d(002) peak as a function of magnetic field to understand the structural transition shown in Fig. 10. We see that a magnetic field in the a-b plane changes the structure along the c-axis. The longer-range structural change onsets near 3 T with increasing magnetic field. The significant hysteresis shows the first-order nature of the transition.

# III. E3. Accessing Magnetic Ground States in $A_2Mo_3O_8$ Systems

A comprehensive study of magnetic exchange interactions was obtained by mapping DFT total energies onto a Heisenberg model [40]. We use the conventions in that work to describe the system. The magnetic order is determined by $J_{\parallel}$ (a-b plane exchange) and $J_{\perp}$ (interlayer exchange) type terms, as labeled in Fig. 1(a). It was found that the balance between the FRM and AMF phases is determined exclusively by the $J_{\perp}$ exchange values. We point out that these parameters are extremely sensitive to the c-axis length. Hence, the tunability of magnetic phases with a single parameter (the c-axis) is expected and observed here. The stiffness of the a-b layer relative to the c-axis makes the magnetic phase tunable by strain via growth on mismatched substrates, uniaxial pressure, or hydrostatic pressure.

The combined structural parameters and spin on the transition metal site produce unique magnetic ground states for the Mn, Fe, Co and Ni systems. Compared to $Fe_2Mo_3O_8$ ($Fe^{2+}$, S=2, L=0), $Mn_2Mo_3O_8$ ($Mn^{2+}$, S=5/2,L=0) orders in the same way as the Fe system (with $T_N \approx 42$ K). The system undergoes a spin flop transition leading to noncollinear canting of the $O_h$ and $T_d$ spins [41]. For $Co_2Mo_3O_8$ ($Co^{2+}$, S=3/2, L=0), the same Fe system AF state is observed with ($T_N \approx 41$ K) at low magnetic fields but with no metamagnetic transition at moderate fields. At a field above 20 T, however, two distinct spin-flop transitions are observed [42]. $Ni_2Mo_3O_8$ ($Ni^{2+}$, S=1, L=0) has a very low magnetic transition temperature of 5.5 K to a noncolinear stripe-type phase [43]. We now show how the atomic structures in these systems are related in terms of lattice parameters.

In terms of the c/a ratio, note from Fig. 2(b) and its inset that the full Zn doping range in $(Fe_{1-y}Zn_y)_2Mo_3O_8$ can be accessed by isotropic pressure between ambient pressure and 10 GPa. Furthermore, if we consider doping by Mg, Mn, Co, and Ni, using published lattice parameters [30,31], we obtain the c/a ratios of 1.717±0.001, 1.772±0.003, 1.719±0.003, and 1.715±0.003, respectively). The star symbols in the main panel of Fig. 2(b), which are labeled Co and Ni, show the corresponding c/a ratio for $Co_2Mo_3O_8$ and $Ni_2Mo_3O_8$, respectively. Hence, for a fixed spin on the transition metal site $Fe_2Mo_3O_8$ ($Fe^{2+}$, S=2, L=0), we can access the structural space of the Co, Mn, and Ni-based systems. We now see that we can use pressure to replace doping in the Mn-doped $Fe_2Mo_3O_8$ system. The same approach applies to the doped $Mn_2Mo_3O_8$, $Co_2Mo_3O_8$, and $Ni_2Mo_3O_8$ systems.

## IV. Summary

The results indicate that the c-axis length is a critical parameter in driving magnetic order and determining the magnetic ground-state phase. Our combined measurements suggest that by tuning the lattice parameter mechanically, we should enable one to raise the magnetic ordering temperature and possibly tune the magnetic field-induced polarization. Specifically, we note that GPa-level isotropic pressures can be applied to pure Fe2Mo3O8 to reproduce the results of Zn doping. The close proximity of the c/a ratio in $Fe_2Mo_3O_8$ of other systems (such as $A_2Mo_3O_8$, A=Co, Mn, and Ni), in terms of GPa-level pressure (1-10 GPa), points to the utility of using pressure to replace chemical doping to access a broad range of magnetic ground states in $A_2Mo_3O_8$ systems. With the Poisson ratio of approximately -1 derived from the hydrostatic measurement results, one may utilize film growth methods or uniaxial pressure to access the structural parameter space. We note that, in a given structural configuration and near the structural phase boundary, one can cross the boundary with very small strain values, as evidenced by the magnetostriction curves in Fig. 9. Hence, this system is an excellent platform for magnetic order switching in films grown on piezoelectric substrates. Magnetic fields modify the potential wells in which the atoms moving (due to charge redistribution). Furthermore, fields applied near a magnetic phase boundary can induce structural distortions.

## V. Acknowledgments

NSF Award DMR-2313456 supports this work. Work at Rutgers University is supported by DOE Grant No. DE-GF02-07ER46382. Diamond cell preparation and testing were conducted at NSLS2 bealine 22-IR-1. Diamond cell preparation, testing and sample loading were conducted at NSLS2 beamline 22-IR-1, a facility supported by the National Science Foundation Division of Earth Sciences (EAR) SEES: Synchrotron Earth and Environmental Science (No. EAR– 2223273) and Chicago/DOE Alliance Center (CDAC) (No. DE-NA-0004153). High-pressure powder XRD measurements were performed at APS beamline 13-BM-D GeoSoilEnviroCARS (The University of Chicago, Sector 13 at Argonne National Laboratory under supported by the National Science Foundation – Earth Sciences via SEES: Synchrotron Earth and Environmental Science (EAR–2223273). Temperature-dependent and magnetic field-dependent X-ray absorption measurements were conducted at the CHESS Synchrotron facility at the Center for High-Energy X-ray Sciences (CHEXS) PIPOXS beamline ID2A at Cornell University. CHEXS is supported by the National Science Foundation (BIO, ENG, and MPS Directorates) under award DMR-2342336. This research used 28-ID-1 of the National Synchrotron Light Source II, the National Synchrotron Light Source II, a U.S. Department of Energy (DOE) Office of Science User Facility operated for the DOE Office of Science by Brookhaven National Laboratory under Contract No. DE-SC0012704. Single-crystal X-ray diffraction measurements were performed at NSF's ChemMatCARS, Sector 15 at the Advanced Photon Source (APS), Argonne National Laboratory (ANL), which is supported by the Divisions of Chemistry (CHE) and Materials Research (DMR), National Science Foundation, under grant number NSF/CHE- 1834750. This research used resources of the Advanced Photon Source, a U.S. Department of Energy (DOE) Office of Science User Facility operated for the DOE Office of Science by Argonne National Laboratory under Contract No. DE-AC02-06CH11357. This research used resources of the National Energy Research Scientific Computing Center (NERSC), a U.S. Department of Energy Office of Science User Facility located at Lawrence Berkeley National Laboratory, operated under Contract No. DE-AC02-05CH11231. No. DE-AC02-05CH11231. We are indebted to Elisabeth Chloe Bodnaruk, Daniel M. Sabol,

Dr. Ernest Fontes, and Dr. Christopher Pollock for the extensive work required to enable the installation, operation, and continued support of the NJIT 10 T superconducting electromagnet system at the CHESS PIPOXS beamline ID2A

## Author Contributions

Diamond cell preparation, testing, and sample loading were conducted by Z.C. and T. A.T. High-pressure X-ray diffraction measurements and analysis were conducted by T.A.T., S.L., S.C., and V.P. X-ray single crystal diffraction measurements and analysis were conducted by T.A.T., T.C. and Y.-S.C. X-ray powder diffraction measurements and analysis were conducted by T.A.T. and M.A. XAFS measurements and analysis were conducted by T.A.T., C. P., and S. A. DFT modeling was conducted by T.A.T. Samples used were prepared by K. W. and S.-W. C. The manuscript was prepared by T. A. T. and M. A.

## Competing Interests

The authors declare no competing interests

## Data availability

Data will be made available on request to the corresponding authors.

**Fig. 1.** (a) Unit cell of hexagonal $Fe_2Mo_3O_8$ showing the $Fe1O_4$ tetrahedra (Fe1 site) site and $Fe2O_6$ octahedra (Fe2 site) with all symmetry unique atoms labeled. A pair of Fe1 and Fe2 sites exist in each unit cell. See structural details in Tables S1 to S3 of the supplementary document. The magnetic exchange pairs $\mathbf{J}_{II}$, $\mathbf{J}_{\perp 1}$, and $\mathbf{J}_{\perp 2}$ are also labeled in the figure. (b) Temperature-dependent atomic displacement ratio of Upara/Uperp ($U_{33}/(U_{11}/2 + U_{22}/2)$) for the O2 site related to the apex of the $Fe1O_4$ tetrahedra, revealing the suppression of this ratio on passing through the magnetic ordering temperature.

**Fig. 2.** (a) Surface plot of the pressure dependence of the powder diffraction pattern for a selected region of angle (see also Fig. S3). (b) Pressure-dependent lattice parameter changes (left y-axis) with c-axis changes given by the blue symbols and the a-axis changes given by the black symbols. The c/a ratio is given by the red symbols (right y-axis). Note that under these hydrostatic conditions, the c-axis compresses more readily than the a-axis. The error bars are smaller than the symbols. The inset shows the c/a ratio (black circles) for the zinc-doped system as a function of the doping parameter y in $(Fe_{1-y}Zn_y)_2Mo_3O_8$ taken from Ref. [31]. The green triangles show the near-constant value of the a-axis with Zn doping. Note the similarity in c/a of the full Zn doping range and the compression data (blue points). The stars in the main figure labeled Co and Ni show the corresponding c/a ratio for $Co_2Mo_3O_8$ and $Ni_2Mo_3O_8$, respectively, with lattice parameters from Ref [31].

**Fig.3.** (a) XAFS structure function (at 0 T and 5 K) showing peaks at R' ≈1.5 Å (average Fe-O bond) and at R'≈ 3.0 A (Fe1-Fe2, Fe-Mo, and Fe-O bonds). R' is not corrected for the scattering and central atom phase shifts, so peaks appear at lower positions than the real distances. (b) Surface plot of the temperature dependence of the peak amplitudes between 5 K and 280 K.

**Fig. 4.** (a) Extracted radial structure function of the Fe-O bonds (both Fe1 and Fe1 sites) for corrected distances with curves for representative temperatures. (b) Surface plot of the full temperature dependence of the Fe-O bonds.

**Fig. 5.** (a) Temperature-dependent $a$ (blue squares) and $c$ (red squares) lattice parameters compared with Debye model fits (dashed lines) to the corresponding data from 189 K to 294 K and extrapolated to low temperatures. For both $a$ and $c$ lattice parameters, the model follows the measured data down to ~ 150 K and then deviates strongly at low temperatures. The c/a ratio (red squares) also changes abruptly near 60 K (see crossing straight lines). (b) Extracted width (blue symbols) and length (red symbols) of d(002) spacing revealing structural rearrangement near 140 K and near the $T_N$.

**Fig. 6.** Atom projected phonon density of states (DOS) showing the total Mo, Fe, and O contribution and the Fe1 and Fe2 separated contributions. The O DOS corresponding to oxygen in tetrahedra (Td) about Fe1 and in octahedra (Oh) about Fe2 are also given.

**Fig. 7.** (a) Fe K-Edge XAFS structure functions (55 K) at 0 T and 1 T (blue solid and blue dashed) and at 8 T and 9 T (red solid and red dashed). The difference between the 0 T and 1 T signals or the 8 T and 9 T signals indicates the level of uncertainty. (b) Surface plot showing the magnetic field dependence of the R' ≈1.5 Å peak amplitude between 0 T and 9 T. The arrows Fe-O polyhedral are robust up to 7 T and then exhibit broadening of their distributions. (c) Surface plot showing the magnetic field dependence of the R' ≈3.0 Å peak amplitude between 0 T and 9 T. The arrows show regions of suppressed correlation above 3 T. All measurements are at 55 K.

**Fig. 8.** (a) Mo K-Edge XAFS structure functions (55 K) at 0 T and 0.5 T (blue solid and blue dashed) and at 7.5 T and 8.5T (red solid and red dashed). (b) Surface plot showing the magnetic field dependence of the Mo-O peak amplitude between 0 T and 8 T. (c) Surface plot showing the magnetic field dependence of the Mo-Mo and Mo-Fe Oh peak amplitude between 0 T and 8 T. All measurements are at 55 K.

**Fig. 9.** (a) Geometry for selective peak analysis used for the d(002) peak at 55 K. From the full pattern (see Fig. S2), the first peak in the powder pattern corresponds to the spacing between (002) planes. With the magnetic field along y, this reflection allows determination of changes in the c-axis when the magnetic field is in the a-b plane. Only single crystal grains, such as those at the center of the orange square, contribute to the ring on the 2D detector. The full ring is generated by grains in this geometry but randomly rotated about the y-axis. Grains rotated about the z-axis also contribute. (b) Detector image of X-ray diffraction pattern. (c) Change in the d(002) length as a function of field for increasing fields up to 5 T (red symbols) and decreasing fields (blue symbols) back to 0 T. The solid lines are second-order polynomial fits to the ramp-up and ramp-down AF region data. The dashed line is for the ramp down FRM region. The FRM region is reached at ≈3.8 T on increasing field, while the field must be reduced to ≈3.0 T to return to the AF phase. This hysteresis is consistent with the first-order nature of the AF to FRM magnetic field-induced transition.

**Fig. 10.** Changes in the width of the $d_{002}$ peak as a function of the magnetic field. Hence, a magnetic field in the a-b plane changes the structure along the c-axis. In terms of the magnetic field regions of change, these results are consistent with the XAFS results. The longer-range structural change onsets near 3 T, increasing the magnetic field.

**Fig. 1.**

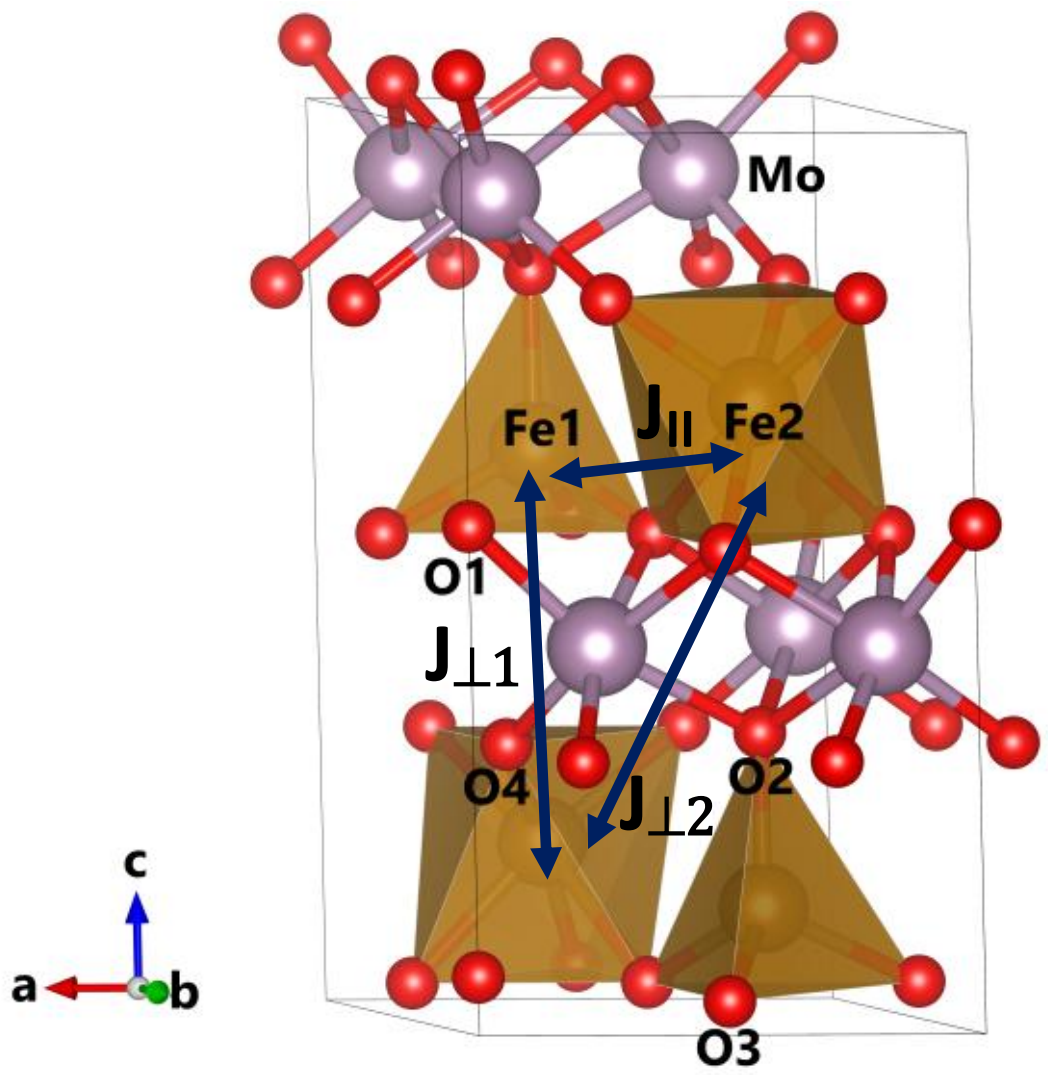

Mo
Fe1
J||
Fe2
O1
J⊥1
O4
J⊥2
O2
O3
c
a
b

(a)

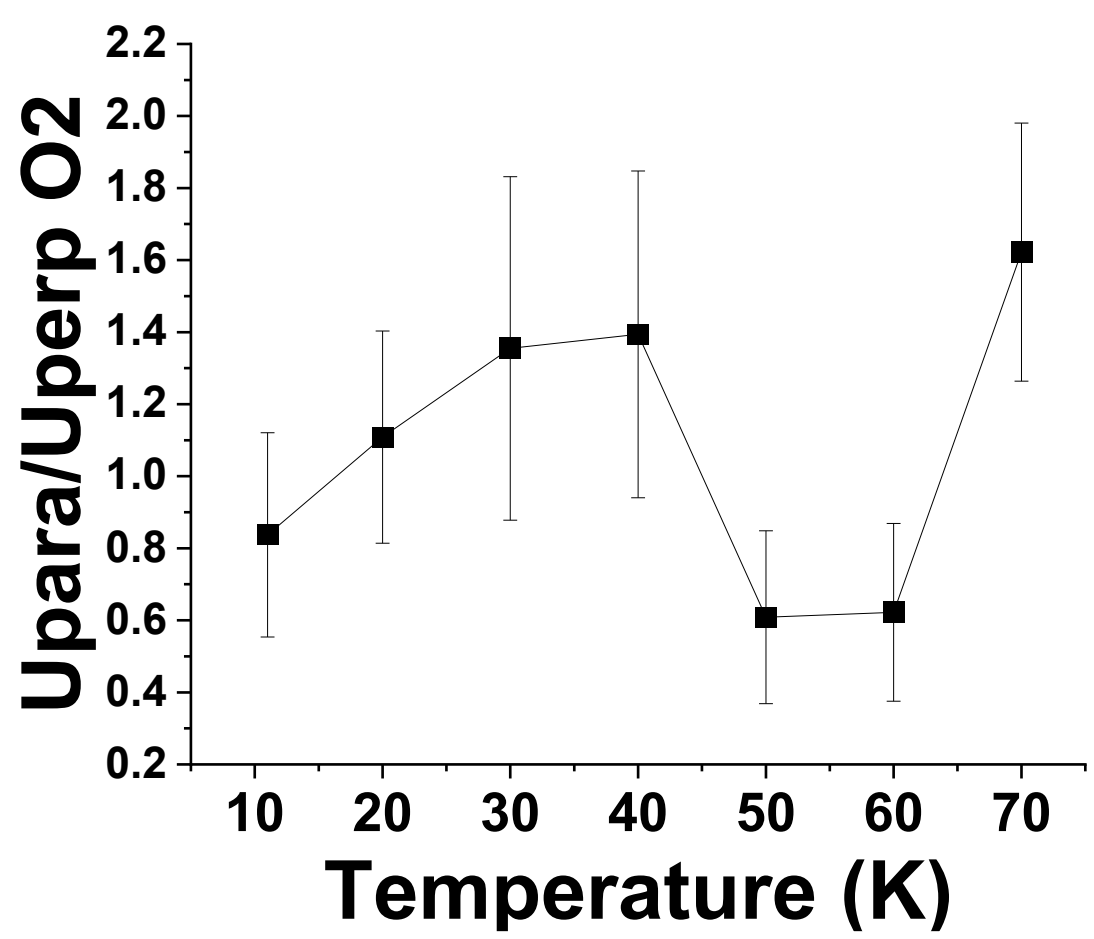

Upara/Uperp O2
Temperature (K)
2.2
2.0
1.8
1.6
1.4
1.2
1.0
0.8
0.6
0.4
0.2
10
20
30
40
50
60
70

(b)

⊥

**Fig. 2.**

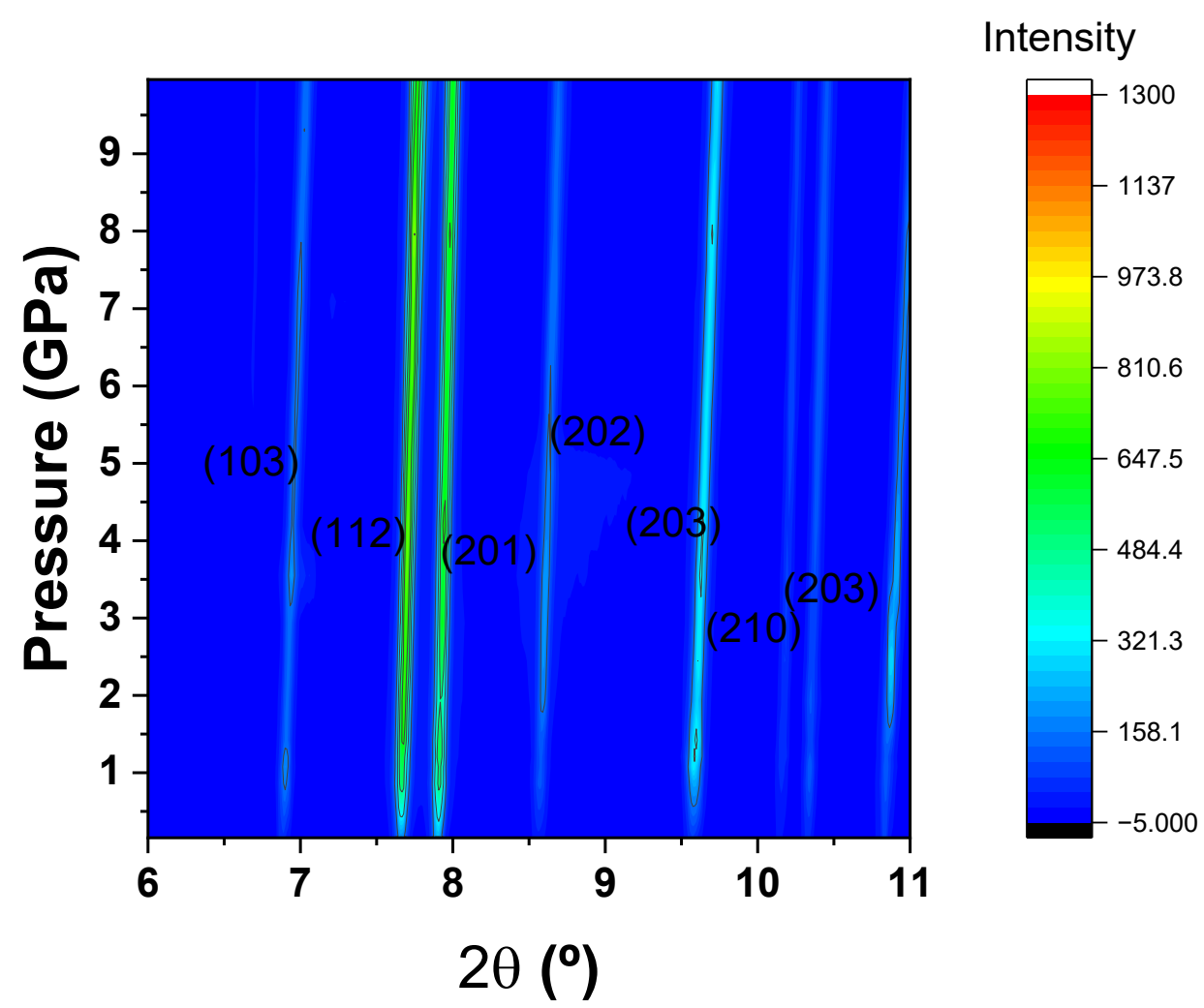

Intensity
1300
1137
973.8
810.6
647.5
484.4
321.3
158.1
−5.000
Pressure (GPa)
(103)
(112)
(201)
(202)
(203)
(210)
(203)
2θ (º)

**(a)**

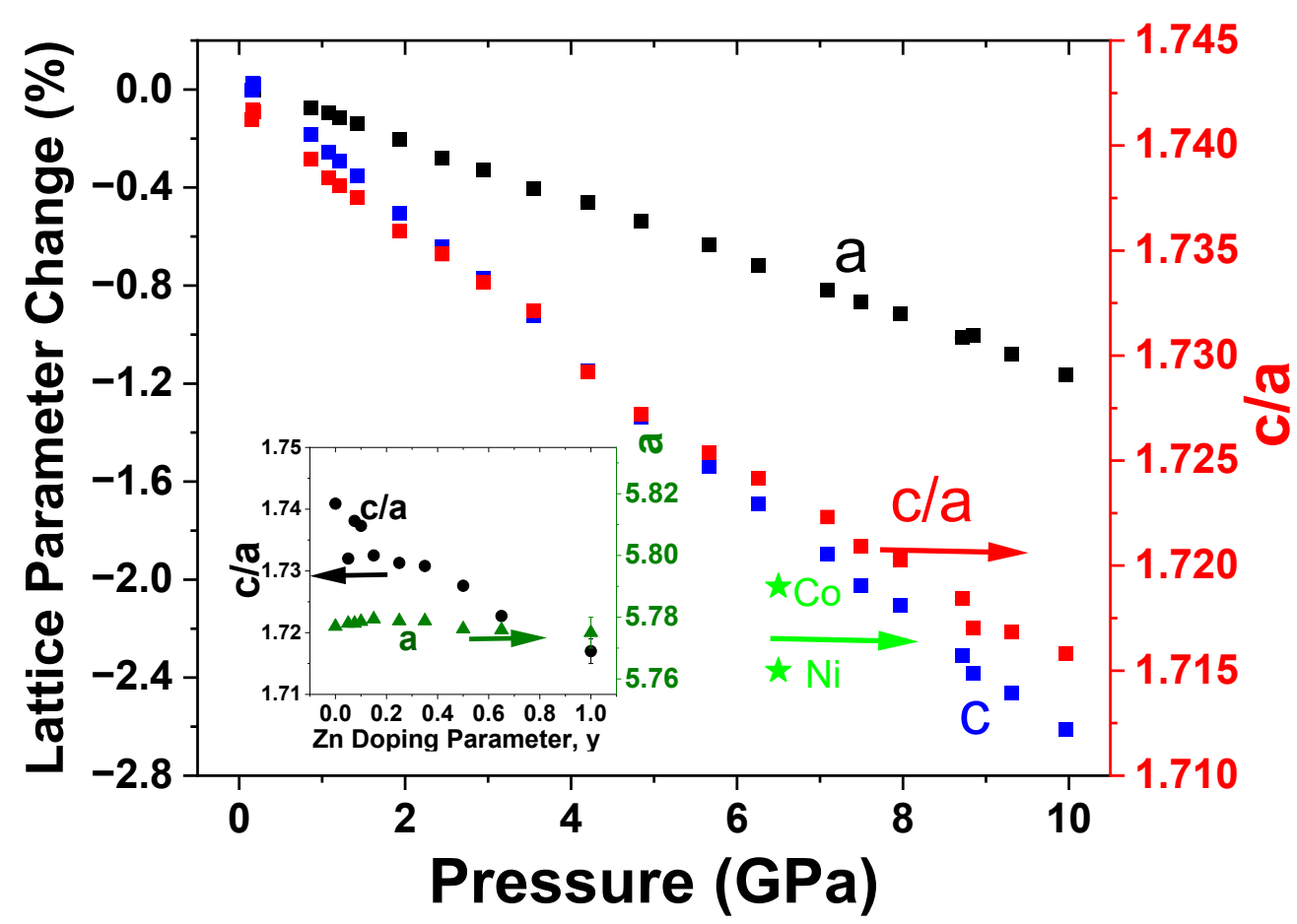

Lattice Parameter Change (%)
Pressure (GPa)
c/a
a
c/a
Co
Ni
c
Zn Doping Parameter, y

**(b)**

**Fig. 3.**

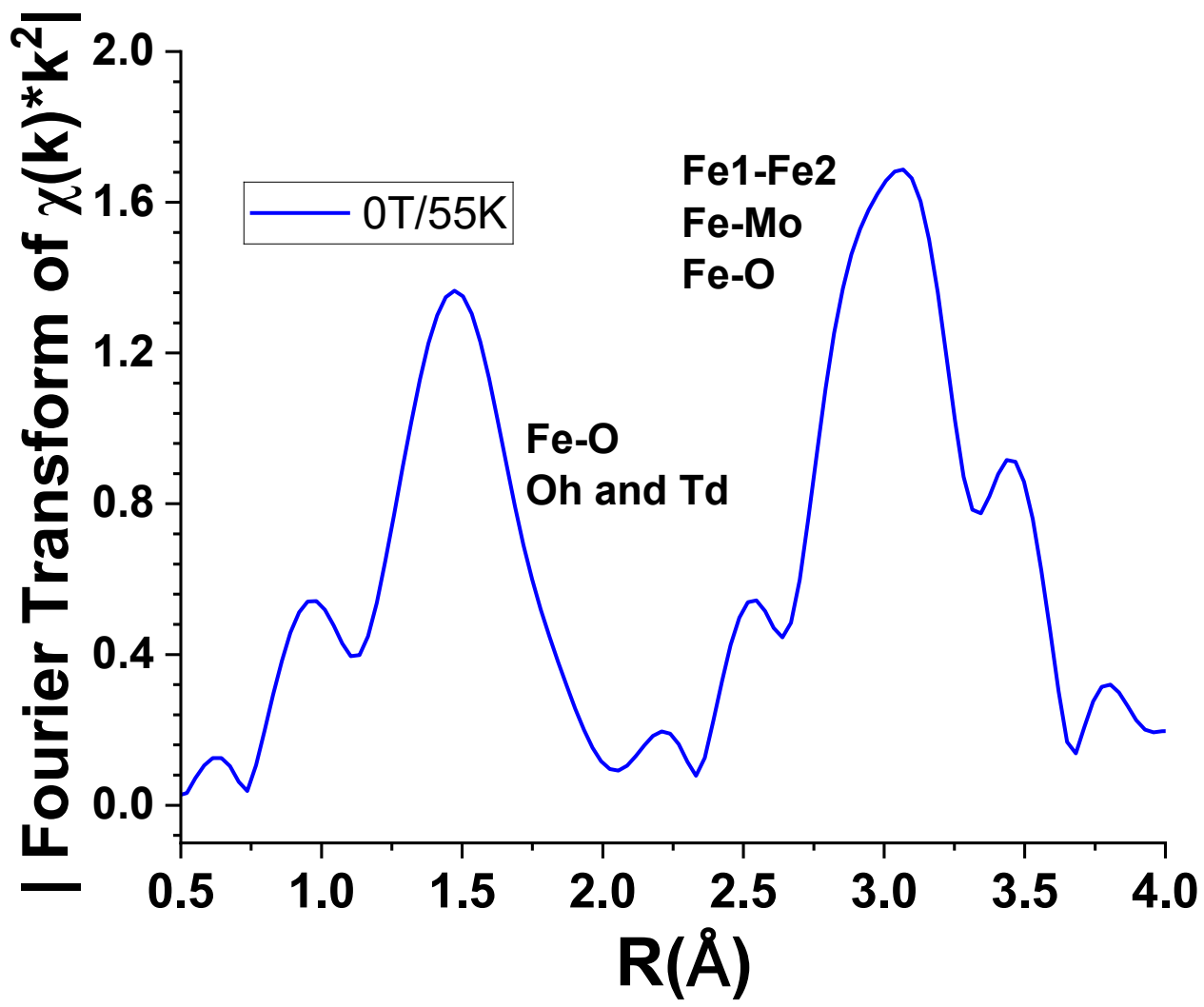

| Fourier Transform of χ(k)*k² |
0T/55K
Fe1-Fe2
Fe-Mo
Fe-O
Fe-O
Oh and Td
R(Å)

**(a)**

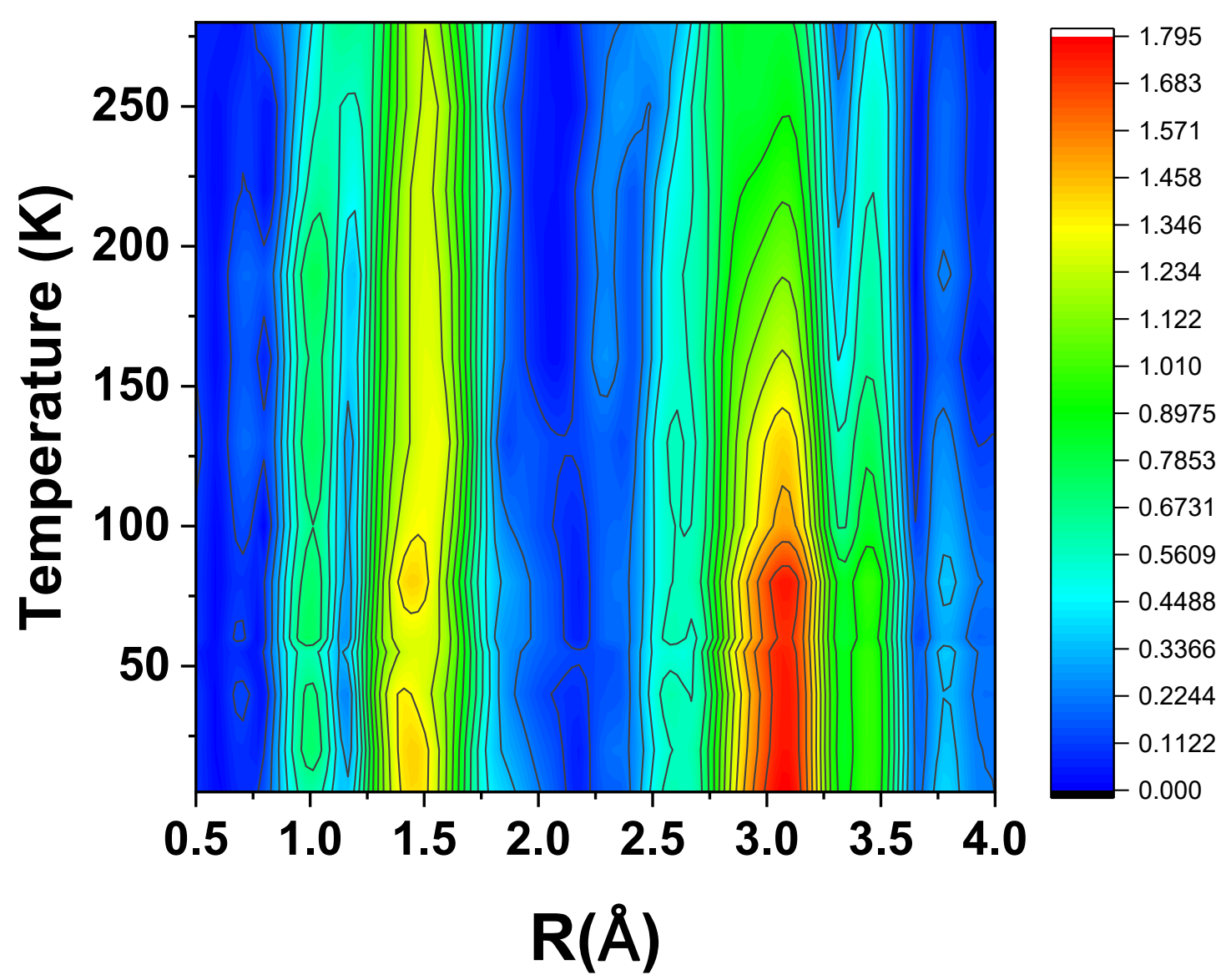

Temperature (K)
R(Å)

**(b)**

**Fig. 4.**

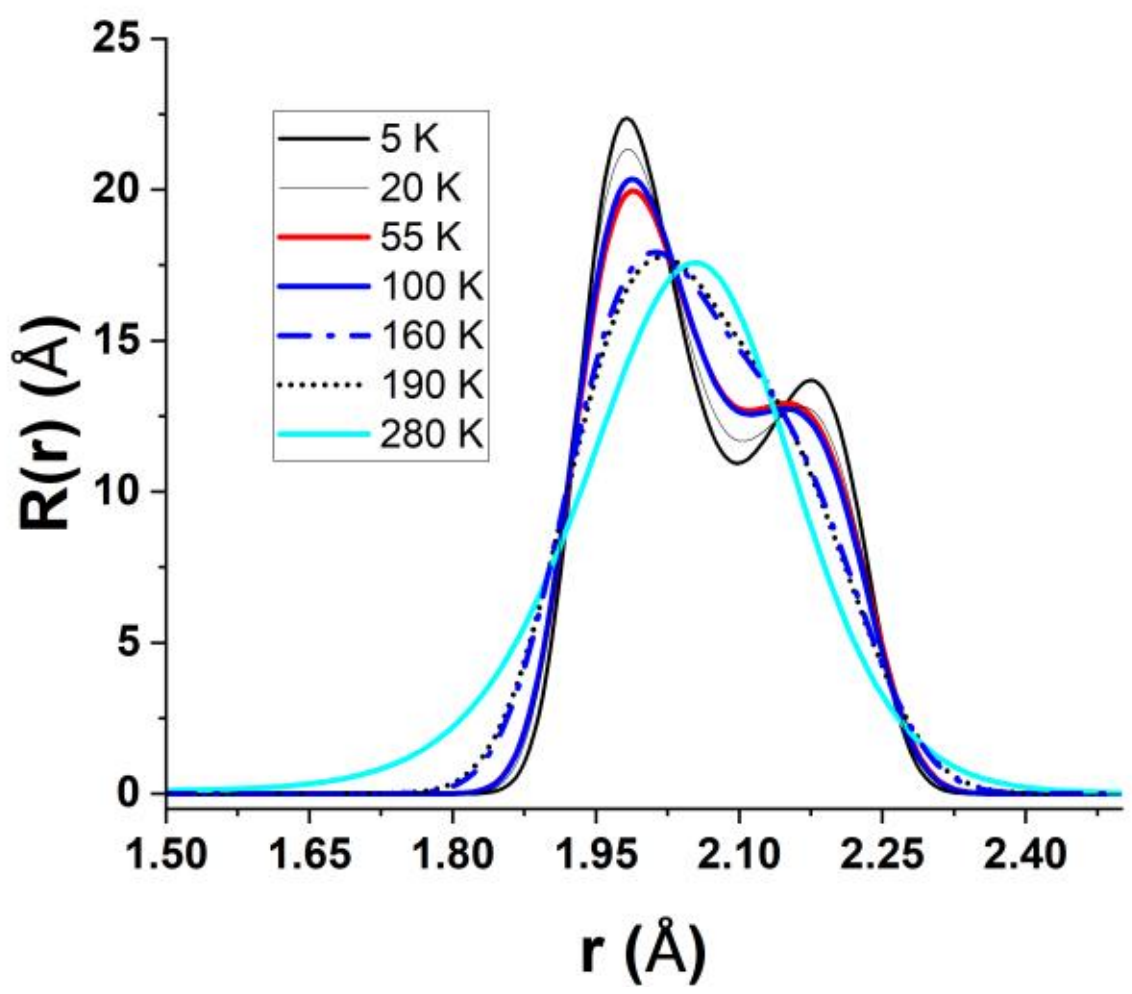

25
20
15
10
5
0
R(r) (Å)
5 K
20 K
55 K
100 K
160 K
190 K
280 K
1.50
1.65
1.80
1.95
2.10
2.25
2.40
r (Å)

**(a)**

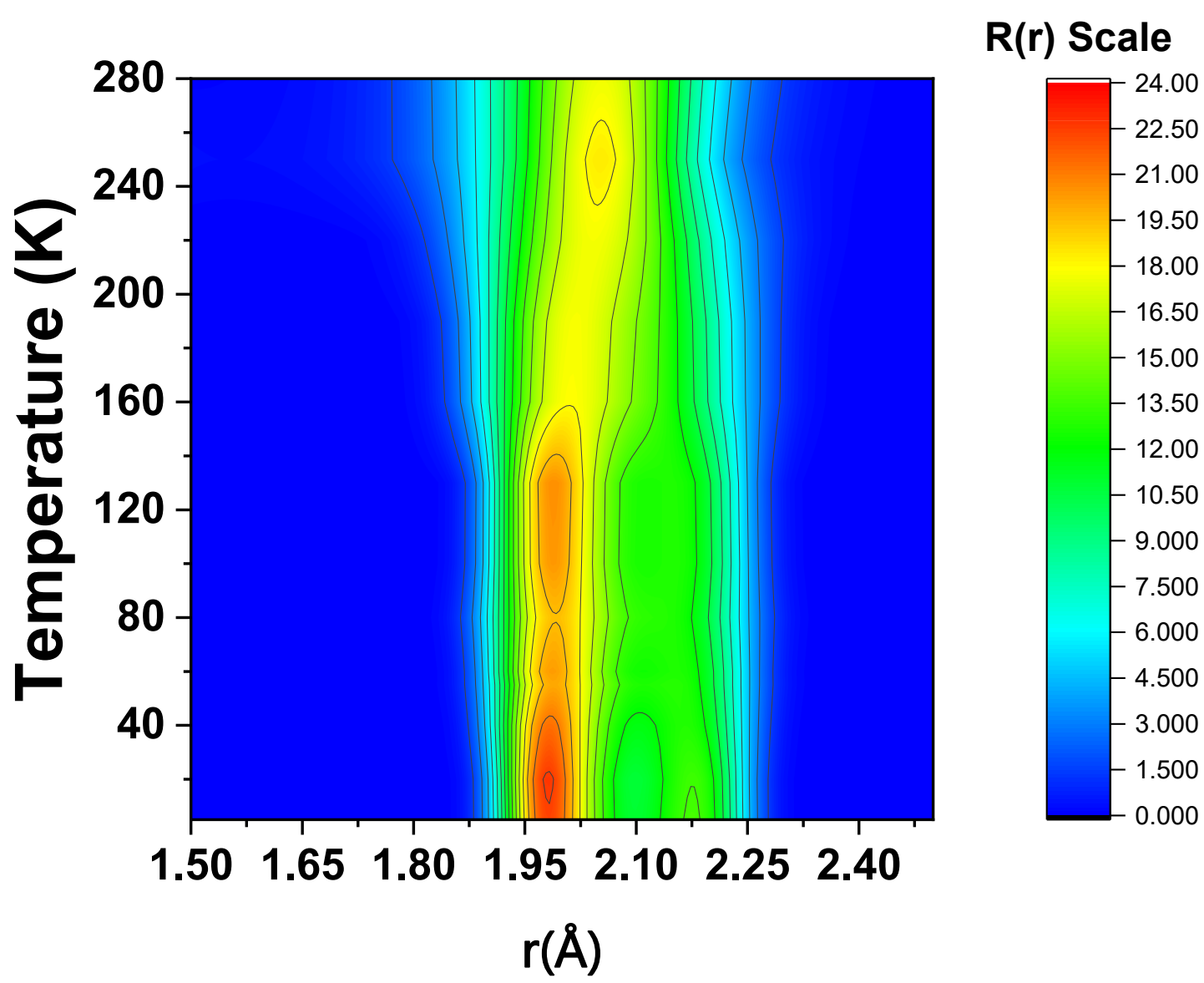

R(r) Scale
24.00
22.50
21.00
19.50
18.00
16.50
15.00
13.50
12.00
10.50
9.000
7.500
6.000
4.500
3.000
1.500
0.000
Temperature (K)
280
240
200
160
120
80
40
1.50
1.65
1.80
1.95
2.10
2.25
2.40
r(Å)

**(b)**

**Fig. 5.**

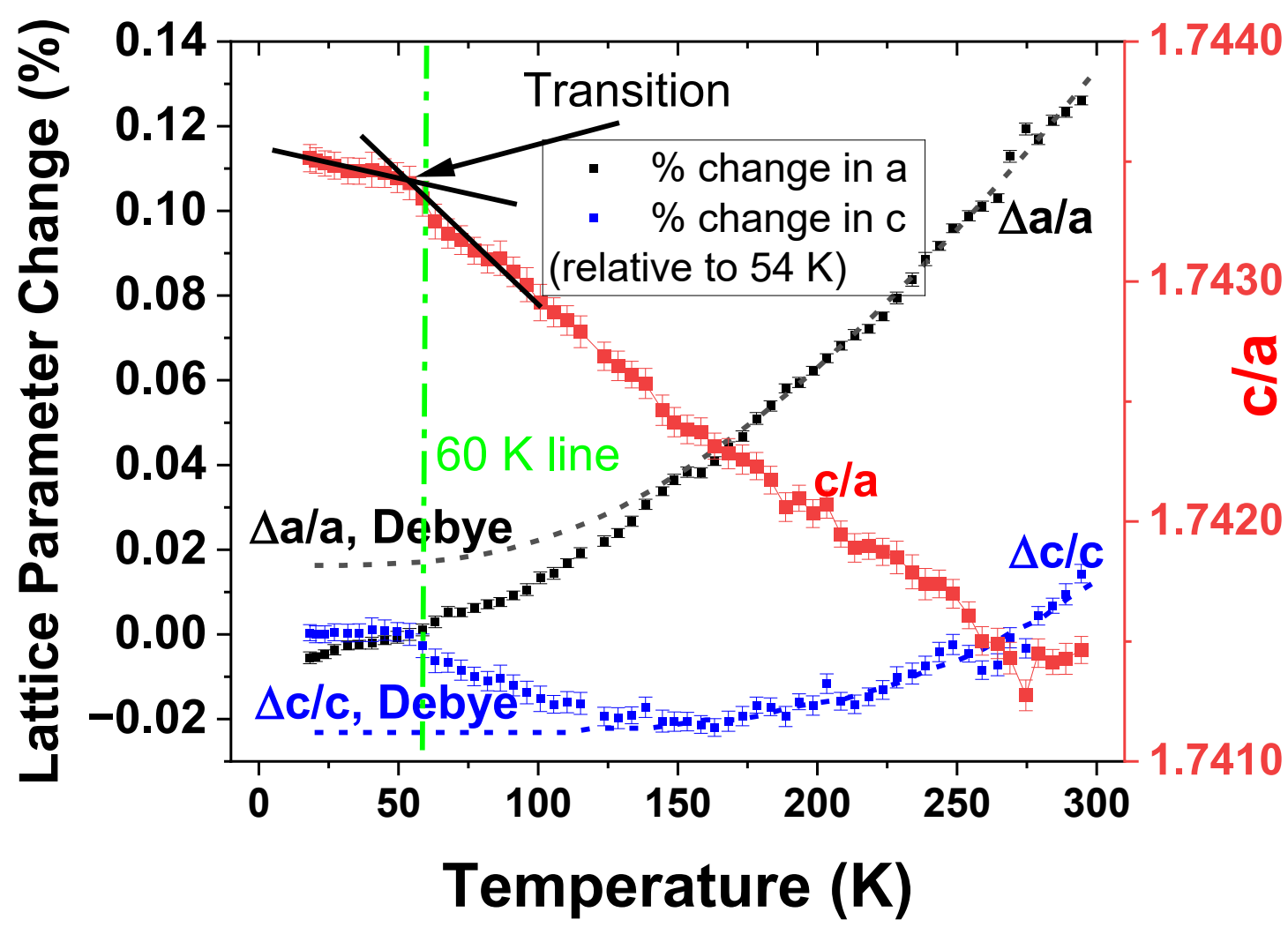
Transition
% change in a
% change in c
(relative to 54 K)
Δa/a
c/a
60 K line
Δa/a, Debye
Δc/c
Δc/c, Debye
Lattice Parameter Change (%)
c/a
Temperature (K)

**(a)**

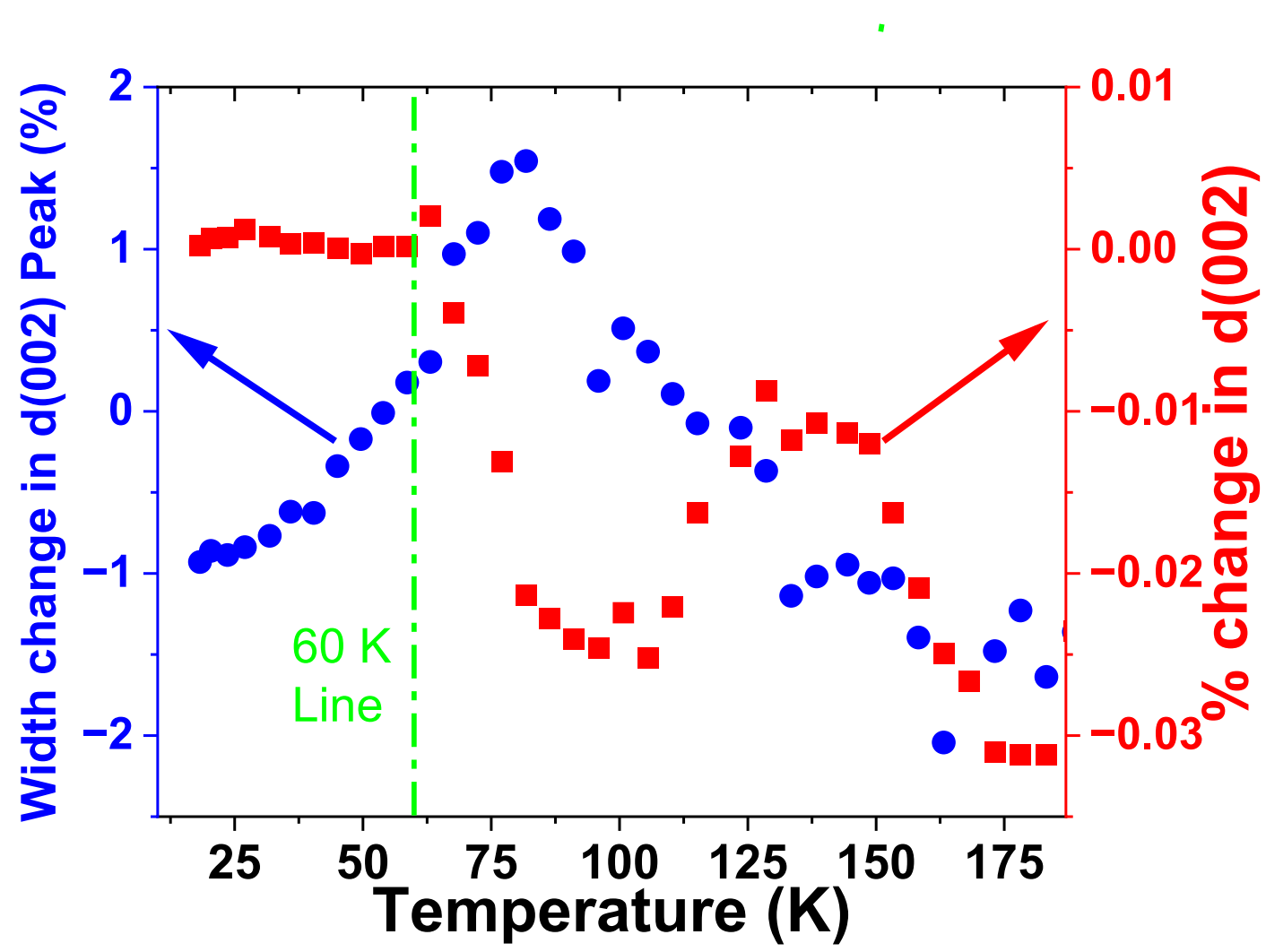
Width change in d(002) Peak (%)
% change in d(002)
60 K
Line
Temperature (K)

**(b)**

**Fig. 6.**

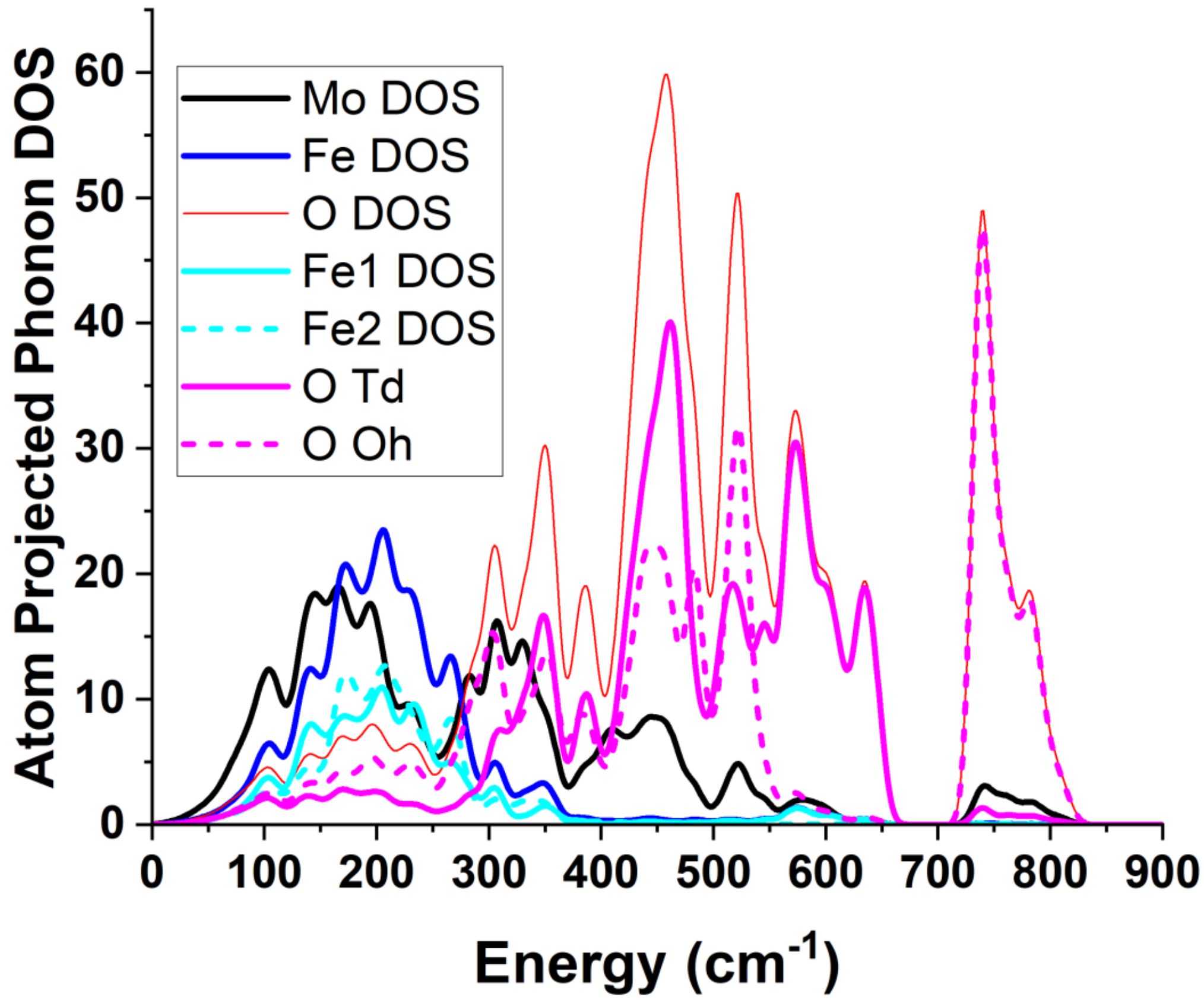
Mo DOS
Fe DOS
O DOS
Fe1 DOS
Fe2 DOS
O Td
O Oh
Atom Projected Phonon DOS
0
10
20
30
40
50
60
0
100
200
300
400
500
600
700
800
900
Energy (cm-1)

**Fig. 7.**

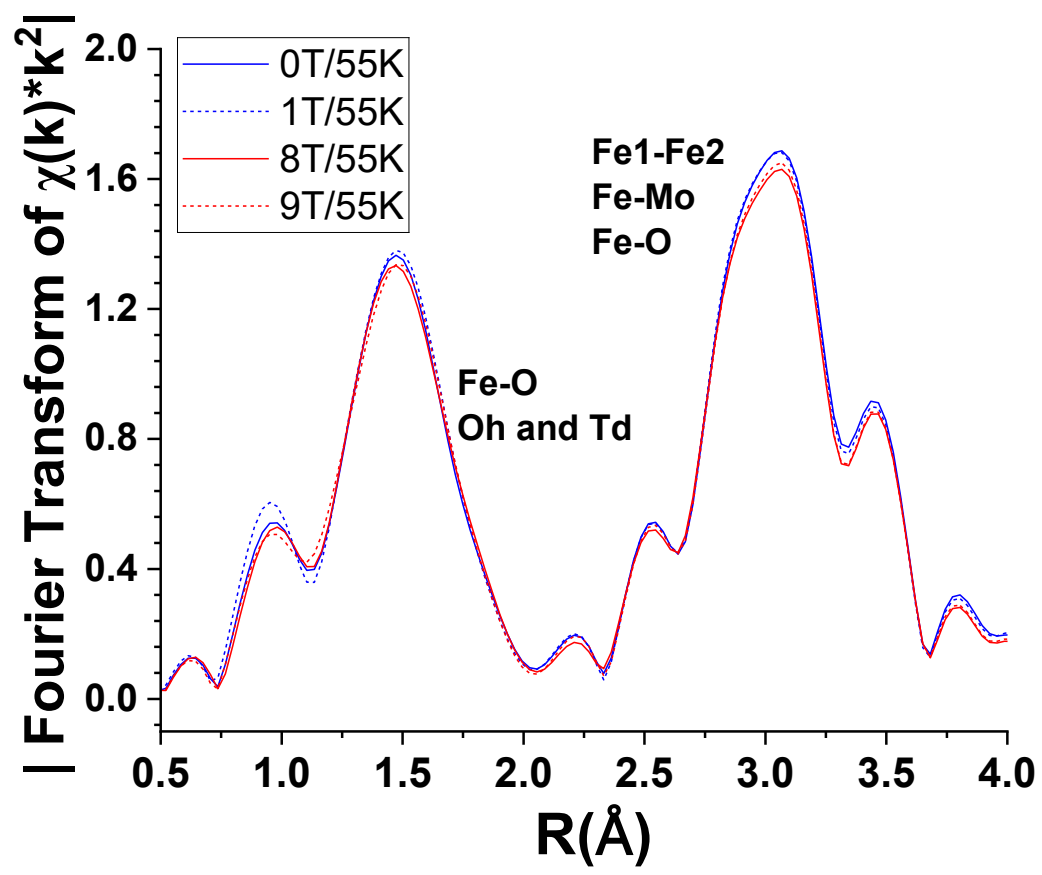

0T/55K
1T/55K
8T/55K
9T/55K
Fe1-Fe2
Fe-Mo
Fe-O
Fe-O
Oh and Td
| Fourier Transform of χ(k)*k²|
R(Å)

**(a)**

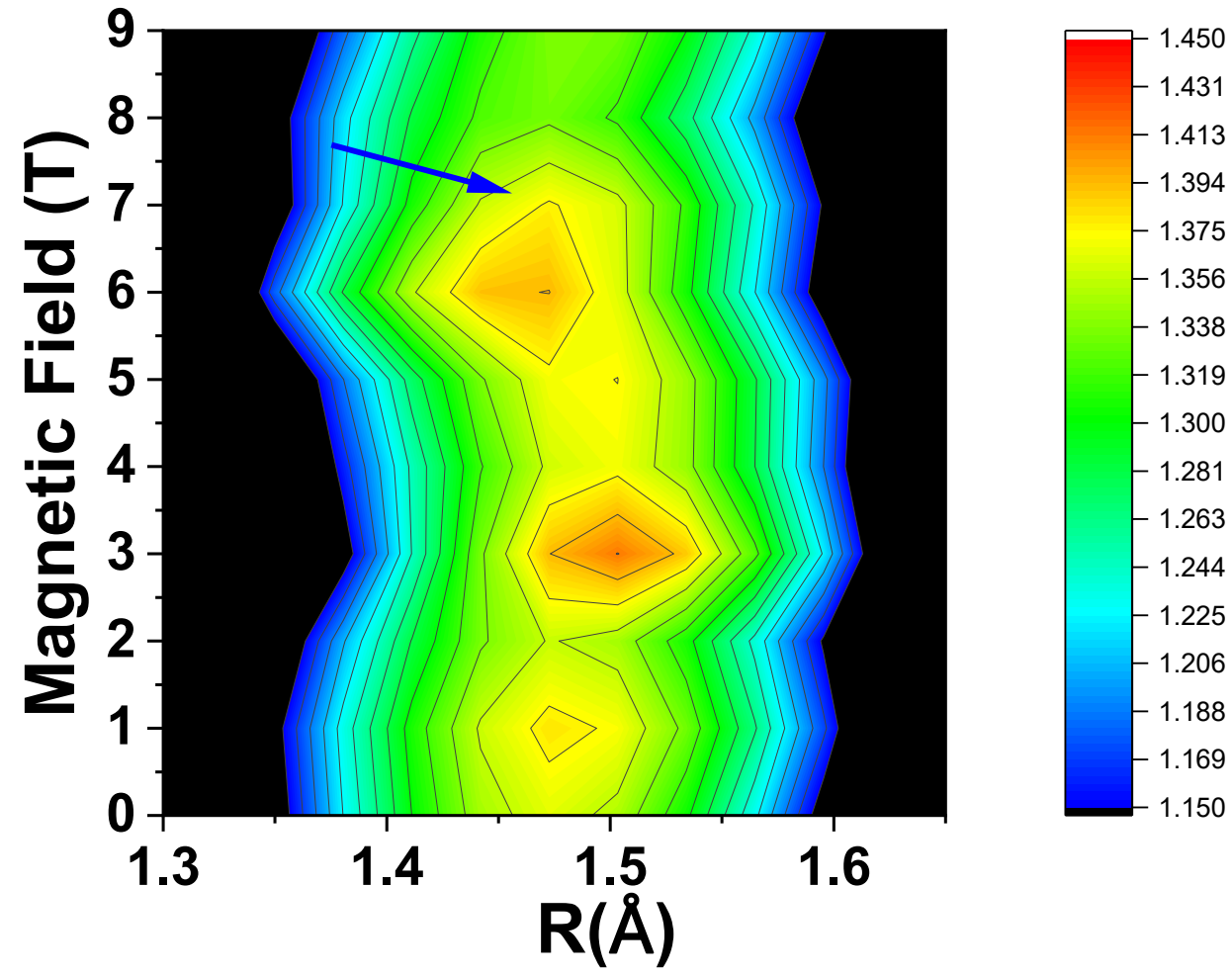

Magnetic Field (T)
R(Å)

**(b)**

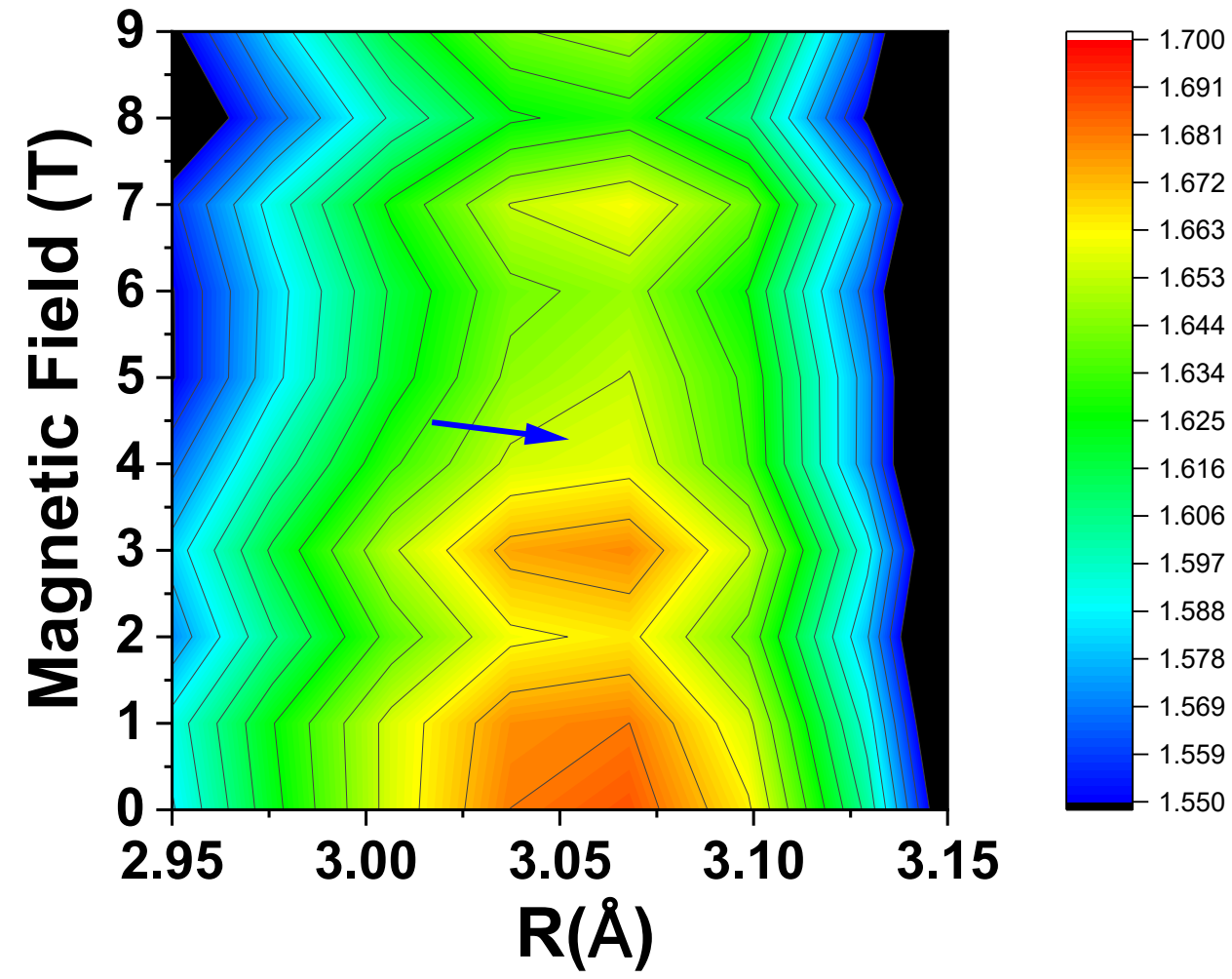

Magnetic Field (T)
R(Å)

**(c)**

Fig. 8.

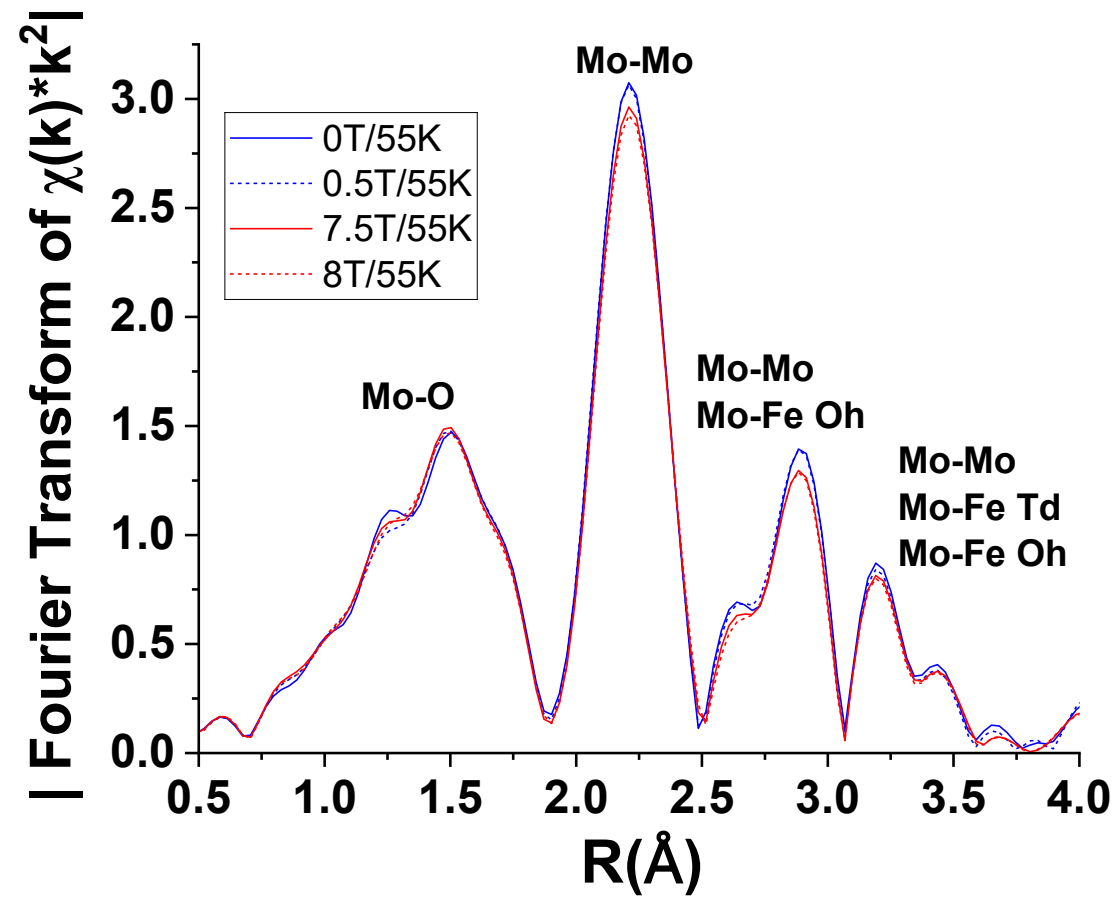

Mo-Mo
0T/55K
0.5T/55K
7.5T/55K
8T/55K
Mo-O
Mo-Mo
Mo-Fe Oh
Mo-Mo
Mo-Fe Td
Mo-Fe Oh
| Fourier Transform of χ(k)*k²|
R(Å)

(a)

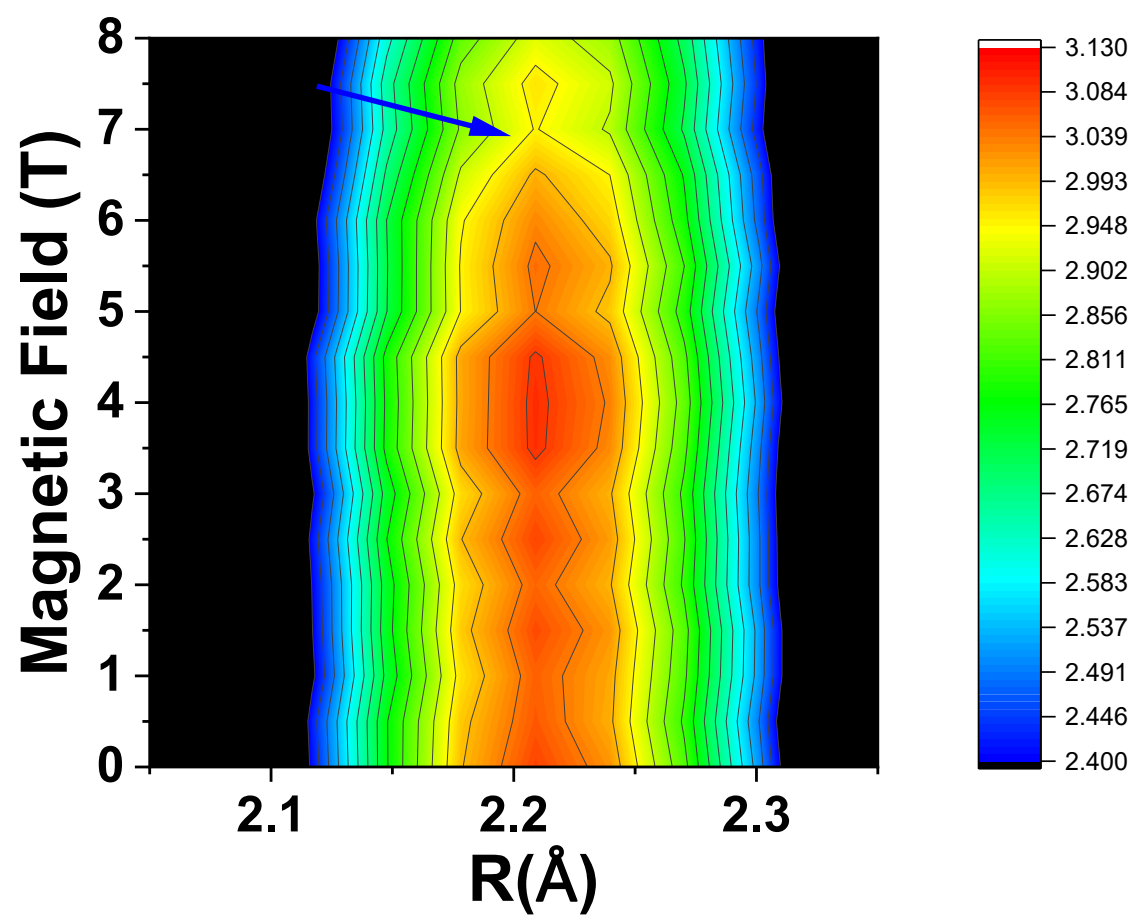

Magnetic Field (T)
R(Å)
3.130
3.084
3.039
2.993
2.948
2.902
2.856
2.811
2.765
2.719
2.674
2.628
2.583
2.537
2.491
2.446
2.400

(b)

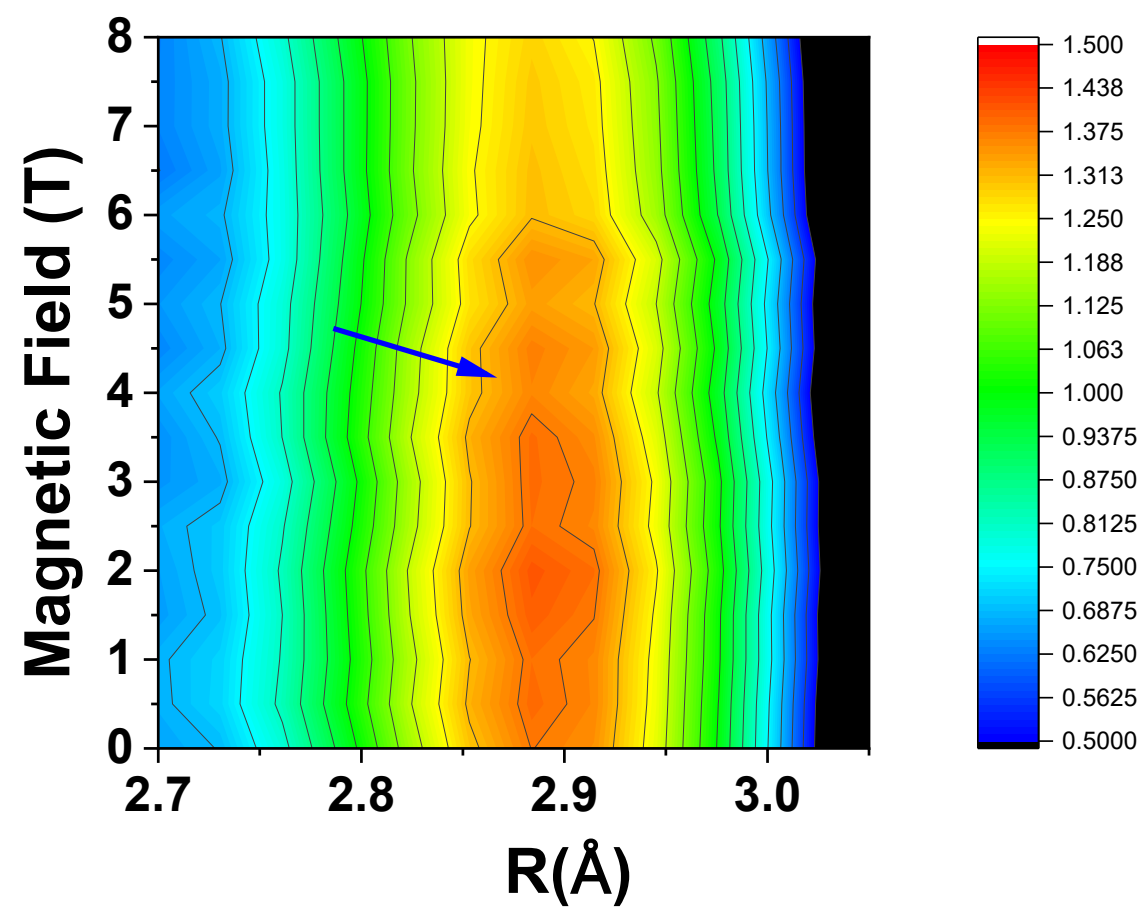

Magnetic Field (T)
R(Å)
1.500
1.438
1.375
1.313
1.250
1.188
1.125
1.063
1.000
0.9375
0.8750
0.8125
0.7500
0.6875
0.6250
0.5625
0.5000

(c)

**Fig. 9.**

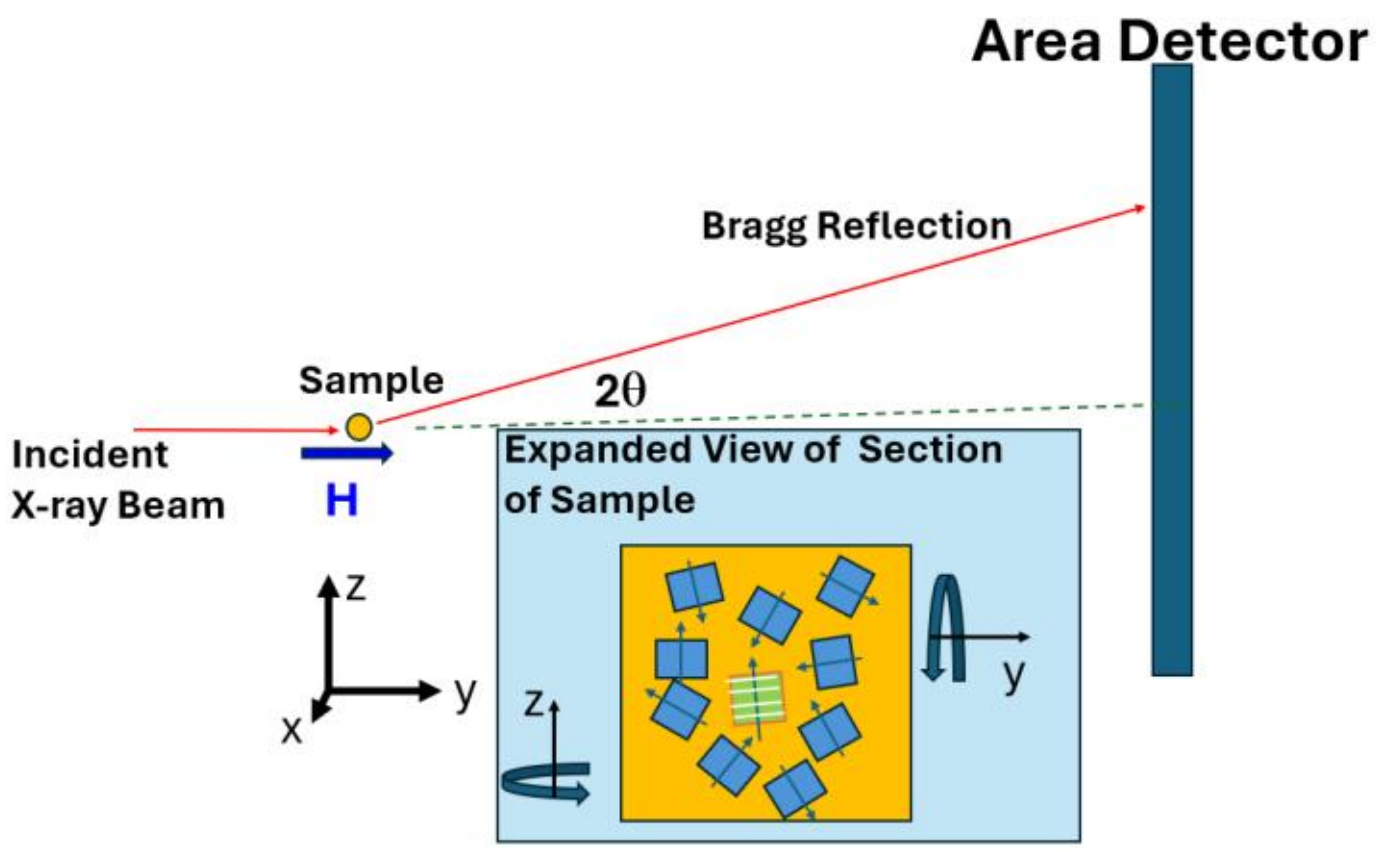

Area Detector
Bragg Reflection
Sample
2θ
Incident
X-ray Beam
H
Expanded View of Section
of Sample
z
y
x
z
y

(a)

(b)

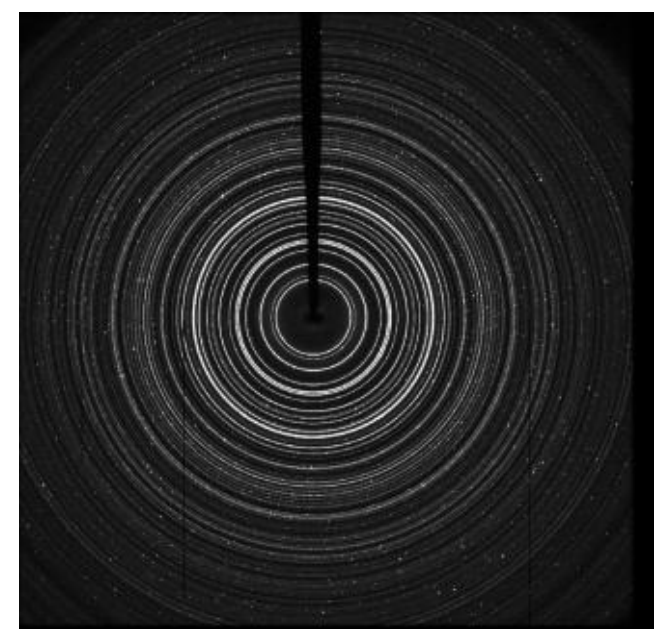

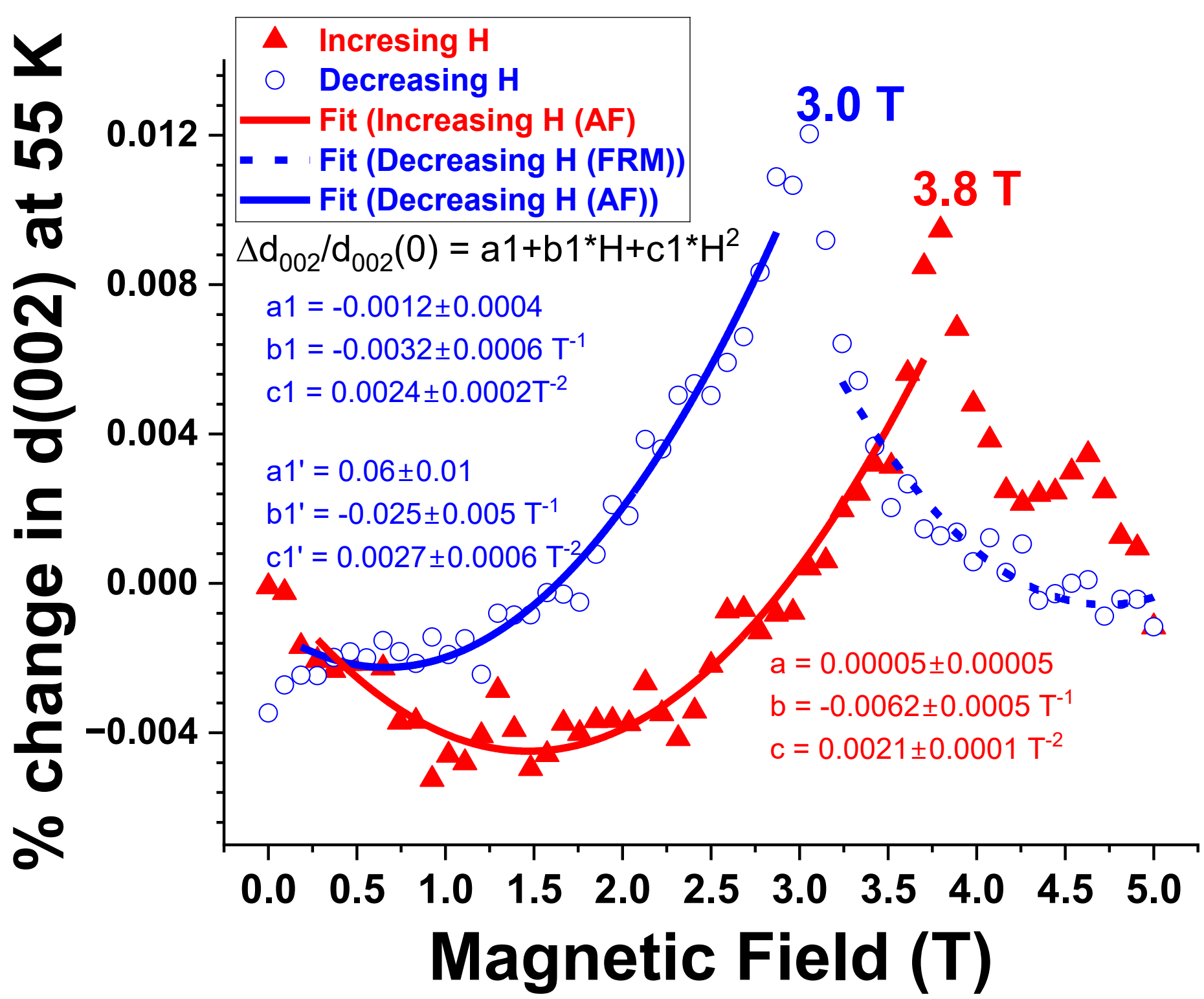

Incresing H
Decreasing H
Fit (Increasing H (AF)
Fit (Decreasing H (FRM))
Fit (Decreasing H (AF))
Δd002/d002(0) = a1+b1*H+c1*H2
a1 = -0.0012±0.0004
b1 = -0.0032±0.0006 T-1
c1 = 0.0024±0.0002T-2
a1' = 0.06±0.01
b1' = -0.025±0.005 T-1
c1' = 0.0027±0.0006 T-2
3.0 T
3.8 T
a = 0.00005±0.00005
b = -0.0062±0.0005 T-1
c = 0.0021±0.0001 T-2
% change in d(002) at 55 K
0.012
0.008
0.004
0.000
−0.004
0.0
0.5
1.0
1.5
2.0
2.5
3.0
3.5
4.0
4.5
5.0
Magnetic Field (T)

(c)

**Fig. 10.**

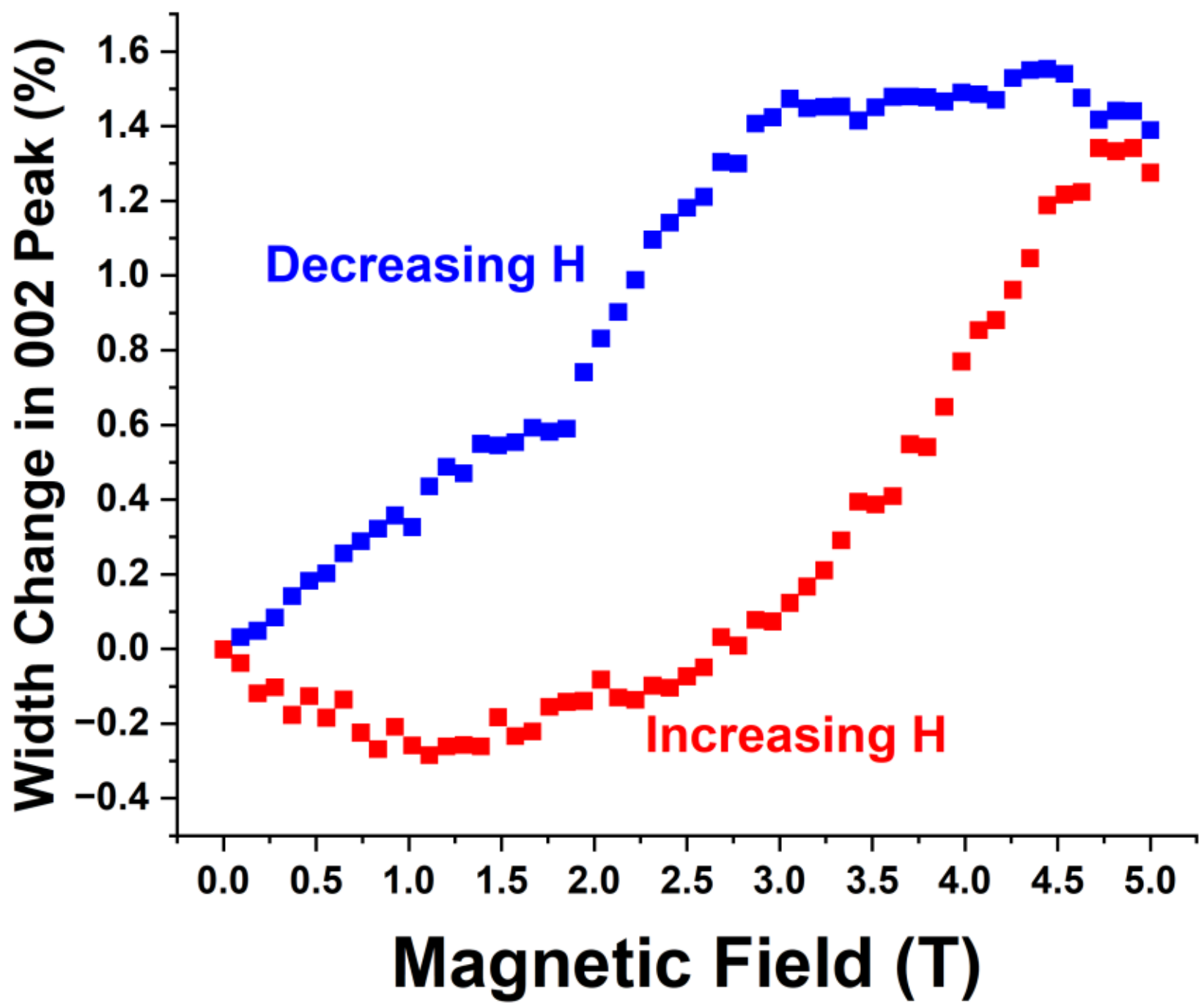

Width Change in 002 Peak (%)
1.6
1.4
1.2
1.0
0.8
0.6
0.4
0.2
0.0
−0.2
−0.4
Decreasing H
Increasing H
0.0
0.5
1.0
1.5
2.0
2.5
3.0
3.5
4.0
4.5
5.0
Magnetic Field (T)

# Critical Structural Parameter Determining Magnetic Phases in the $Fe_2Mo_3O_8$ Altermagnet System

T. A. Tyson[1,3,*], S. Liu[1], S. Amarasinghe[1], K. Wang[2,3],
S. Chariton[4], V. Prakapenka[4], T. Chang[4], Y.-S. Chen[4], Z. Liu[5]
C. J. Pollock[6], S.-W. Cheong[2,3], and M. Abeykoon[7,*]

[1]Department of Physics, New Jersey Institute of Technology, Newark, NJ 07102, USA
[2]Department of Physics and Astronomy, Rutgers University, Piscataway, NJ 08854, USA
[3]Rutgers Center for Emergent Materials, Rutgers University, Piscataway, NJ 08854, USA
[4]Center for Advanced Radiation Sources, The University of Chicago, Argonne, IL 60439, USA
[5]Department of Physics, University of Illinois at Chicago, Illinois 60607-7059, USA
[6]Cornell High Energy Synchrotron Source, Wilson Laboratory, Cornell University, Ithaca, New York 14853, USA
[7]National Synchrotron Light Source II, Brookhaven National Laboratory, Upton, NY 11973, USA

# (Supplementary Information Document)

*Corresponding Authors:

T. A Tyson, e-mail: tyson@njit.edu

M. Abeykoon, e-mail: aabeykoon@bnl.gov

## Sample Preparation

Crystals of $Fe_2Mo_3O_8$ were prepared by chemical vapor transport at 1000 ºC via 10-day run followed by furnace cooling [1]. The ambient pressure powder x-ray diffraction (XRD) and x-ray absorption fine-structure (XAFS) measurements were based on materials derived by grinding and then sieving the crystal materials using a 500-mesh sieve (25 μm particle size). Diamond cell preparation, testing and sample loading were conducted at NSLS2 beamline 22-IR-1. For the high-pressure diffraction measurements, the 500 mesh powders were reground to produce materials with reduced particle size (≈ 1 micron ) and pressed into ≈10 μm thick sheets using polished steel cubes.

## Synchrotron-Based X-Ray Single Crystal Diffraction Measurements

Diffraction measurements were conducted on a 18μm × 10 μm × 10 μm single crystal for measurements at and below 70 K (and on a 30 μm × 25 μm × 20 μm above 70 K) at the Advanced Photon Source (APS) beamline 15-ID-D at Argonne National Laboratory using a wavelength of 0.41328 Å (30 keV). The data were collected with a PILATUS 1M CdTe detector (by DECTRIS, maximum count rate = $10^7$ cps/pixel, counter depth = 20 bit). The large dynamic range enabled the detection of scattering from both weak (O) and strong scatterers (Mo and Fe). An Oxford cryostat system was used to cool the sample in liquid helium vapor. Data were collected between 70 K and 11 K. (A Nitrogen vapor system was used above 70 K). The sample was mounted at the tip of a glass fiber with epoxy and rotated by 360º in steps of 0.3º, yielding 1200 images (1 sec exposure per image). At each temperature, data were collected at two Kappa values corresponding to full sphere data. The distance from the samples to the detector was 130 mm (110 mm at 300K). The image data were processed using APEX3 (Bruker, 2016) [2]. The solution and refinement of the data were done using the program Olex2 [3] after the reflections were corrected for absorption using SADABS. Anomalous scattering corrections (*f*' and *f*'') were induced for all atoms. No assumptions were made about the symmetry in the data collection process. Hence, full spheres in reciprocal space were collected. This resulted in more than 9 times as many reflections with amplitude $F(hkl)|$ as needed for complete data sets based on symmetry. The space group with the optimal fit parameters was found by systematically exploring hexagonal space groups to find the high symmetry space group with the best-fit characterization parameters ($R_1$, $wR_2$, and *GOF*) and no systematic absence violations. Here $R_1$, $wR_2$, and *GOF* (Goodness of fit) are defined as $R_1 = \sum||F_O| - |F_c||/\sum|F_O|$, $wR_2 = \{\sum w(F_O^2 - F_c^2)^2 / \sum w(F_O^2)^2\}^{\frac{1}{2}}$, $GOF = \{\sum w(F_O^2 - F_c^2)^2)/(n-p)\}^{\frac{1}{2}}$, respectively. In these expressions, $F_O$ is the measured scattering amplitudes, $F_c$ is the calculated scattering amplitudes, $w$ is the weight, $n$ is the number of reflections, and $p$ is the total number of fitting parameters. For a good fit to the data, both the $R_1$ and residual charge should be as small as possible, and the *GOF* should approach 1**. For all temperatures studied (300 K and 70 K to 11 K), the best solution was obtained for the P$6_3$mc space group.** The detailed structural data at 11 K, 70 K, and 300 K are given in tables S1, S2, and S3.

## Ambient Pressure Powder Diffraction Measurements for Varying Temperature and Magnetic Field Dependent Powder XRD

Powder diffraction data were collected using Beamline 28-ID-1 (PDF) at NSLS2 (National Synchrotron Light Source), at Brookhaven National Laboratory, with a wavelength of 0.1665 Å in a 1 mm Kapton capillary sealed with clay. Measurements utilized a Perkin Elmer Area detector with a sample-to-detector distance of 750.8 mm. The exact detector-to-sample distance was derived by fits to a Ni powder calibration standard. The measurements were conducted in a continuous-flow liquid helium Cryo Industries cryostat with a temperature sensor next to the sample. This cryostat was inserted into a superconducting Cryo Industries magnet (not in persistent mode). The magnetic field was varied between 0 T and 5 T and then back down to 0 T in steps of 0.1 T. This ramping was done for multiple fixed sample temperatures: 51 K, 55 K, 58 K, 83 K, and 119 K. For these measurements, Rietveld fits were used to extract structural changes with temperature and magnetic field.

## High-Pressure Powder Diffraction Measurements

High-pressure powder XRD measurements at room temperature were performed at APS beamline 13-BM-D (GESCARS) at Argonne National Laboratory. The beam size used was 4 μm (V) x 9 μm (H) with a wavelength of 0.3344 Å. To apply pressure, a standard cylindrical diamond anvil cell with 400-micron culets was used. A rhenium gasket with 200 μm thickness was pre-indented to ~ 43 μm (with 150 μm holes for the sample chamber). Neon was used as the pressure-transmitting medium, and Ruby balls and Au foils were placed near the sample for pressure calibration. The Ruby balls were used for the initial calibration and Ne gas loading. The full range of pressure was calibrated using Au standard diffraction patterns taken at each pressure. The cell was rocked to randomize the sampling of grains in the x-ray beam. The detector-to-sample distance was 234.2 mm (based on a NIST $LaB_6$ standard). A PILATUS 1M detector with 172 μm x 172 μm CdTe-based pixels, was used for measurements. Pressure calibration by a Au standard follows the methods in Ref. [4]. The ambient pressure pattern was recovered on pressure release. The Dioptas program [5] was utilized to integrate the two-dimensional diffraction images (powder rings). More complete details, including the calibration procedure, can be found in Ref. [6]. The lattice parameters at each pressure were extracted by LeBail fits. This model is insensitive to preferred orientation which is possible with the small sample volumes used in diamond anvil cells.

## XAFS Measurements and Data Analysis

Temperature-dependent and magnetic field-dependent x-ray absorption measurements were conducted at the CHESS PIPOXS beamline ID2A at Cornell University. The Fe K-edge measurements were conducted in transmission mode by mixing the 500 mesh powder with BN (https://www-ssrl.slac.stanford.edu/smbin/mucalwebnew.pl) and pressing the powder into 13 mm diameter pellets with a die. This produced a sample with a two attenuation length thickness above the Fe K-edge. The temperature and magnet field were accessed using a variable temperature insert in the NJIT 10T Superconducting Oxford Spectromag4000-10 Magnet System (https://web.njit.edu/~tyson/Supmagnet.html), which is now stationed at ID2A at CHESS. Example results from earlier experiments at NSLS can be found in [7,8].

The incident x-ray energy was selected using a cryogenically cooled Si(111) monochromator and delivered to the sample without focusing, resulting in a 1 x 2 mm spot. Beam intensity was measured before and after the sample using $N_2$-filled ion chambers. At fixed magnetic field and temperature, more than eight scans were taken at each temperature to obtain a high signal-to-noise ratio needed to explore small changes in structure. The temperature uncertainty is < 0.5 K. A Fe foil reference was used for energy calibration. The reduction of the x-ray absorption fine-structure (XAFS) data was performed using standard procedures [9]. Data were collected in monochromator "flyscanning" mode between 6952 eV and 8247eV for the Fe K-edge. Approximately eight 6-minute scans were averaged at each temperature and magnetic field point.

To treat the atomic distribution functions on equal footing at all temperatures, the spectra were modeled in R-space by optimizing the integral of the product of the radial distribution functions and theoretical spectra with respect to the measured spectra. The experimental temperature-dependent spectra were modeled by, $\chi(k) = \int \chi_{th}(k,r)\, 4\pi r^2 g(r) dr$ where $\chi_{th}$ is the theoretical spectrum [10] and g(r) is the real space radial distribution function based on a sum of Gaussian functions ($\chi(k)$ is the measured spectrum) [11] at each temperature as in Ref. (as in Ref. [12]). The atomic distribution g(r) corresponding to first peak in the Mn K-edge XAFS data (Fig. 3(a)) was modeled as a sum of Gaussians representing a distribution with fixed coordination number. Distributions which minimize the difference between the experimental and the model at each temperature are shown in Fig. 4 (as done in our work in Ref. [12]). The k-range 1.4 $Å^{-1}$ < k < 13.0 $Å^{-1}$ and the R-range 0.80 Å < R < 2.23 Å were utilized. The same approach was adopted for Mo K-edge data reduction. Data were collected in monochromator "flyscanning" mode between 19,805 eV and 21585 eV. The k-range 2.78 $Å^{-1}$ < k < 18.1 $Å^{-1}$ was used for Fourier transform signals shown in this work. Approximately six 3-minute scans were averaged at each temperature and magnetic field point for the Mo K-edge data to yield the spectra as shown in Fig. S3(a).

## Computation of the Phonon Density of States

To determine the force constants and phonon DOS for $Fe_2Mo_3O_8$, density-functional calculations in the projector-augmented-wave approach were performed using the VASP code [13]. In the LDA+U approach, U=5 eV and J=1 eV were utilized. Full structural optimization was conducted for both lattice parameters and atomic positions. The LDA exchange functional (Ceperly and Alder as parameterized by Perdew and Zunger [14]) was used to obtain the relaxed structure. The ground-state structure was optimized so that forces on each atom were below 2 x $10^{-5}$ eV/Å. Calculations for a 2×2×2 supercell with a gamma-centered *k*-space grid were utilized. The force constants were calculated by density-functional-perturbation theory with the VASP code. The code Phonopy was used to determine the phonon density of states [15]. Gaussian broadening with a full-width at half maximum of 7.1 $cm^{-1}$ was applied to each phonon DOS spectrum shown in the figures.

**Table S1. Structural Parameters at 11 K***

| Atoms | x(×10⁴) | y(×10⁴) | z(×10⁴) | Ueq (Å²)×10³ |
|---|---|---|---|---|
| Mo | 7080.2(2) | 8540.1(2) | 4250 | 2.57(4) |
| Fe1 | 6666.67 | 3333.33 | 6277.7(6) | 3.66(8) |
| Fe2 | 3333.33 | 6666.67 | 6872.4(6) | 3.17(8) |
| O1 | 10000 | 10000 | 5647(2) | 3.9(4) |
| O2 | 6666.67 | 3333.33 | 8224(3) | 4.0(4) |
| O3 | 4879(2) | 5121(2) | 5378(2) | 4.2(2) |
| O4 | 8335(2) | 6669(3) | 3097(2) | 4.4(2) |

| | $U_{11}$(Å²)×10³ | $U_{22}$(Å²)×10³ | $U_{33}$(Å²)×10³ | $U_{23}$(Å²)×10³ | $U_{13}$(Å²)×10³ | $U_{12}$(Å²)×10³ |
|---|---|---|---|---|---|---|
| Mo | 2.40(5) | 2.60(5) | 2.65(6) | 0.00(2) | -0.01(5) | 1.20(3) |
| Fe1 | 4.1(1) | 4.01(1) | 2.8(2) | 0 | 0 | 2.05(6) |
| Fe2 | 3.4(1) | 3.4(1) | 2.8(2) | 0 | 0 | 1.67(6) |
| O1 | 4.2(5) | 4.2(5) | 3.4(8) | 0 | 0 | 2.1(3 |
| O2 | 4.3(5) | 4.3(5) | 3.5(9) | 0 | 0 | 2.2(3) |
| O3 | 4.2(4) | 4.2(4) | 4.0(6) | 0.1(2) | -0.1(2) | 1.9(4) |
| O4 | 4.5(4) | 4.7(5) | 4.0(6) | -0.5(5) | -0.3(2) | 2.4(2) |

Space Group: $P6_3mc$ (Z=2)
a = 5.77290(14) Å, c = 10.0675(3) Å, Dx = 6.029 g/cm³
V = 290.854(17) Å³
Measurement Temperature: 11.0 K
Crystal Dimensions: 18 μm x 10 μm x 10 μm
Wavelength: λ = 0.41328 Å
Absorption Coefficient: 10.342 mm⁻¹
F(000) = 484.0
2θ Range : 4.706° to 47.206°
-11≤ h ≤9, -9≤ k ≤ 11, and -19 ≤ l ≤16
Number of Measured Reflections: 6624
Flack parameter: 0.06(12)
Independent reflections/restraints/parameters: 676/0/31
Max and Min Peak in Final Difference Map: 0.97(Mo)/-0.51(O2) e-/ Å³
$R_1$ = 0.95 %, $wR_2$ = 2.08.0 %, Goodness of Fit = 1.05(I>=2σ (I))
$R_1$ = 0.95 %, $wR_2$ = 2.08% (all data)
$R_{int}$ = 3.05 %, $R_{sigma}$= 2.02 %

*z-coordinate of Mo is fixed.

**Table S2. Structural Parameters at 70 K***

| Atoms | x(×10⁴) | y(×10⁴) | z(×10⁴) | Ueq (Å²)×10³ |
|---|---|---|---|---|
| Mo | 7079.8(2) | 8539.9(2) | 4250 | 3.08(4) |
| Fe1 | 6666.67 | 3333.33 | 6276.6(7) | 4.56(9) |
| Fe2 | 3333.33 | 6666.67 | 6876.3(6) | 3.87(9) |
| O1 | 10000 | 10000 | 5649(3) | 4.9(4) |
| O2 | 6666.67 | 3333.33 | 8222(3) | 4.7(4) |
| O3 | 4879(2) | 5121(2) | 5377(2) | 5.1(3) |
| O4 | 8333.6(18) | 6667(4) | 3097(3) | 5.1(3) |

| | $U_{11}$(Å$^2$)×10$^3$ | $U_{22}$(Å$^2$)×10$^3$ | $U_{33}$(Å$^2$)×10$^3$ | $U_{23}$(Å$^2$)×10$^3$ | $U_{13}$(Å$^2$)×10$^3$ | $U_{12}$(Å$^2$)×10$^3$ |
|---|---|---|---|---|---|---|
| Mo | 2.72(6) | 2.97(5) | 3.47(7) | 0.01(3) | 0.02(5) | 1.36(3) |
| Fe1 | 4.9(1) | 4.92(12) | 3.8(2) | 0 | 0 | 2.46(6) |
| Fe2 | 3.8(1) | 3.89(12) | 3.8(2) | 0 | 0 | 1.95(6) |
| O1 | 5.5(6) | 5.5(6) | 3.7(9) | 0 | 0 | 2.7(3) |
| O2 | 4.6(6) | 4.6(6) | 5(1) | 0 | 0 | 2.3(3) |
| O3 | 4.2(4) | 4.2(4) | 6.1(8) | 0.3(3) | -0.3(3) | 1.5(5) |
| O4 | 5.2(4) | 4.8(5) | 5.2(8) | -1.2(6) | -0.6(3) | 2.4(3) |

Space Group: $P6_3mc$ (Z=2)
a = 5.77530(14) Å, c = 10.0692(3) Å, Dx = 6.023 g/cm$^3$
V = 290.854(17) Å$^3$
Measurement Temperature: 70.0 K
Crystal Dimensions: 18 μm x 10 μm x 10 μm
Wavelength: λ = 0.41328 Å
Absorption Coefficient: 10.332 mm$^{-1}$
F(000) = 484.0
2θ Range : 4.704° to 46.92°
$-10 \leq h \leq 9$, $-9 \leq k \leq 10$, and $-18 \leq l \leq 16$
Number of Measured Reflections: 5971
Flack parameter: 0.06(14)
Independent reflections/restraints/parameters: 656/0/31
Max and Min Peak in Final Difference Map: 0.48(O1)/-0.50(O1) e-/ Å$^3$
$R_1$ = 0.97 %, $wR_2$ = 2.24.0 %, Goodness of Fit = 1.08(I>=2σ (I))
$R_1$ = 0.97 %, $wR_2$ = 2.24% (all data)
$R_{int}$ = 3.28 %, $R_{sigma}$= 2.20 %

*z-coordinate of Mo is fixed.

**Table S3. Structural Parameters at 300 K***

| Atoms | x(×10[4]) | y(×10[4]) | z(×10[4]) | Ueq (Å[2])×10[3] |
|---|---|---|---|---|
| Mo | 7079.4(2) | 8539.7(2) | 4250 | 3.72(3) |
| Fe1 | 6666.67 | 3333.33 | 6273.4(7) | 6.97(8) |
| Fe2 | 3333.33 | 6666.67 | 6883.1(6) | 5.69(7) |
| O1 | 10000 | 10000 | 5650(3) | 5.6(3) |
| O2 | 6666.67 | 3333.33 | 8221(3) | 5.4(3) |
| O3 | 4881(2) | 5119(2) | 5381(2) | 6.0(2) |
| O4 | 8338(2) | 6676(4) | 3096(2) | 6.6(2) |

| | $U_{11}$(Å$^2$)×10$^3$ | $U_{22}$(Å$^2$)×10$^3$ | $U_{33}$(Å$^2$)×10$^3$ | $U_{23}$(Å$^2$)×10$^3$ | $U_{13}$(Å$^2$)×10$^3$ | $U_{12}$(Å$^2$)×10$^3$ |
|---|---|---|---|---|---|---|
| Mo | 2.91(5) | 3.68(4) | 4.30(5) | -0.07(2) | -0.15(4) | 1.46(2) |
| Fe1 | 7.8(1) | 7.8(1) | 5.3(2) | 0 | 0 | 3.90(6) |
| Fe2 | 6.3(1) | 6.3(1) | 4.6(2) | 0 | 0 | 3.12(5) |
| O1 | 6.3(5) | 6.3(5) | 4.4(7) | 0 | 0 | 3.1(3) |
| O2 | 5.3(5) | 5.3(5) | 5.6(8) | 0 | 0 | 2.6(2) |
| O3 | 5.1(4) | 5.1(4) | 7.1(6) | 0.87(19) | -0.9(2) | 2.0(4) |
| O4 | 6.2(4) | 6.4(5) | 7.2(6) | -1.8(5) | -0.9(2) | 3.2(2) |

Space Group: $P6_3mc$ (Z=2)
a = 5.78200(10) Å, c = 10.0700(3) Å, Dx = 6.009 g/cm$^3$
V = 290.854(17) Å$^3$
Measurement Temperature: 300.0 K
Crystal Dimensions: 30 μm x 25 μm x 20 μm
Wavelength: λ = 0.41328 Å
Absorption Coefficient: 10.307 mm$^{-1}$
F(000) = 484.0
2θ Range : 4.704° to 55.168°
-10≤ h ≤11, -12≤ k ≤ 10, and -13 ≤ l ≤21
Number of Measured Reflections: 8542
Flack parameter: 0.92(2)
Independent reflections/restraints/parameters: 905/0/31
Max and Min Peak in Final Difference Map: 1.08(O2)/-1.48(Fe1) e-/ Å$^3$
$R_1$ = 1.62 %, $wR_2$ =3.22.0 %, Goodness of Fit = 1.10(I>=2σ (I))
$R_1$ = 1.64 %, $wR_2$ = 3.22% (all data)
$R_{int}$ =4.82 %, $R_{sigma}$= 2.93 %

*z-coordinate of Mo is fixed.

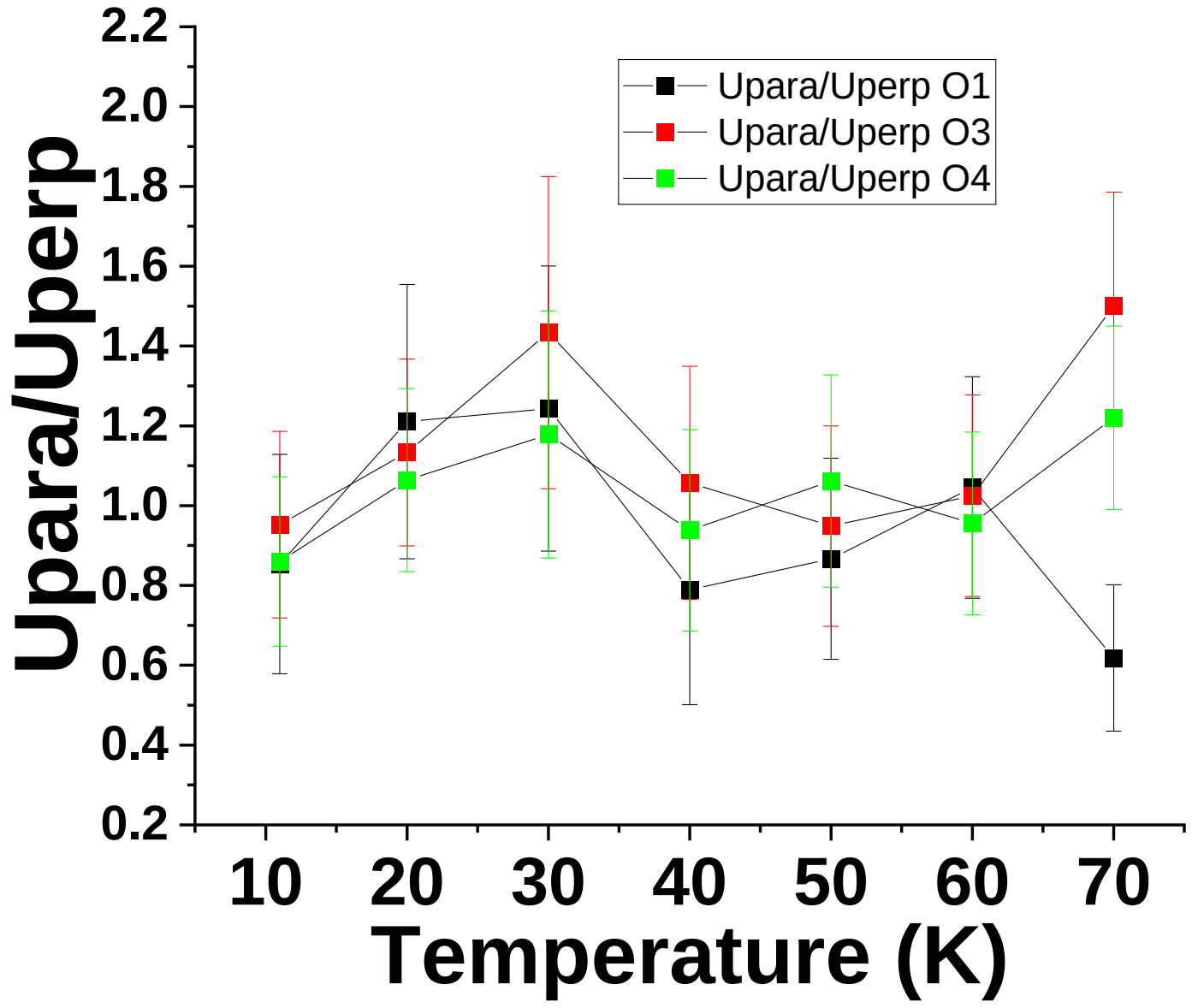

(a)

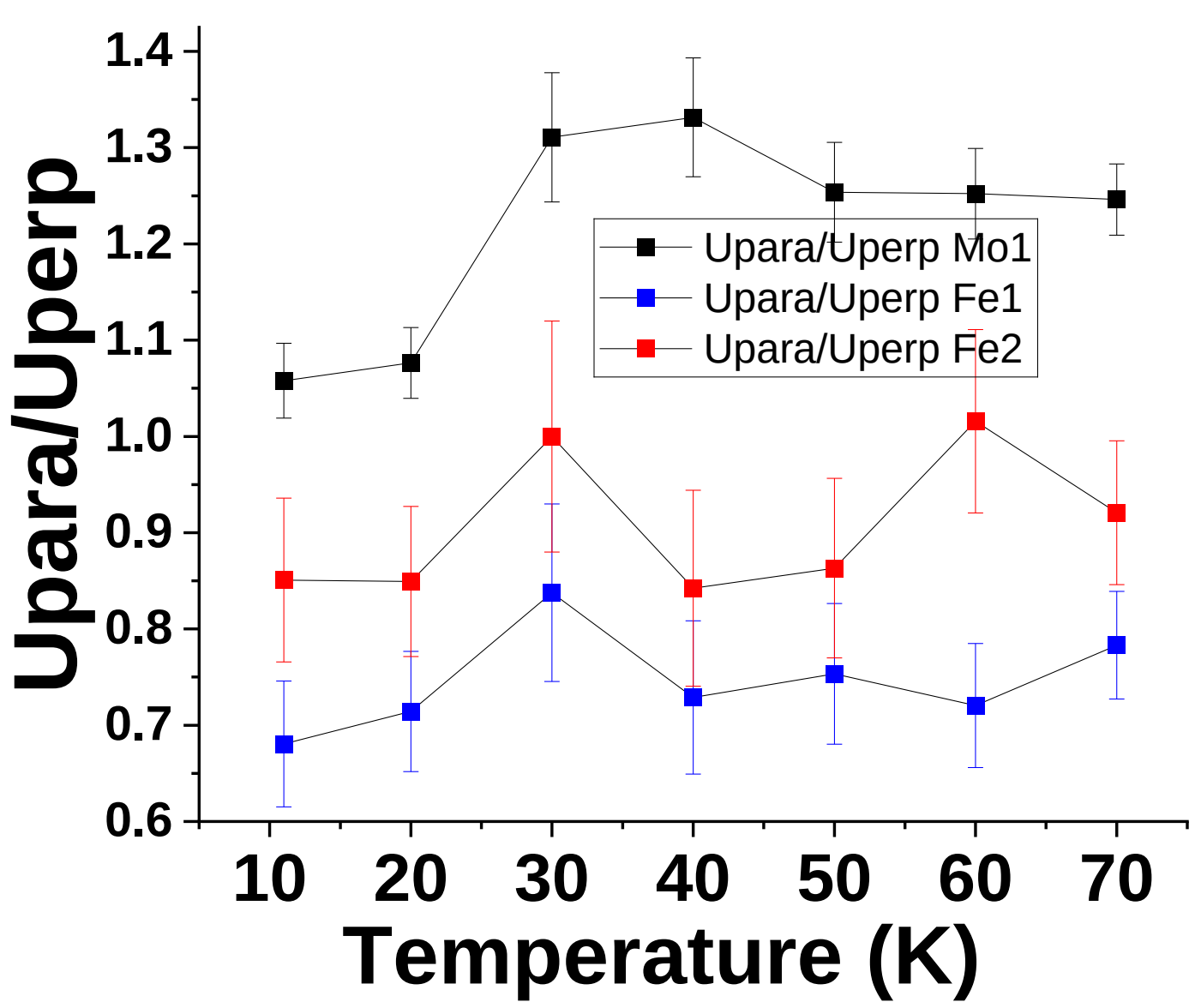

(b)

**Fig. S1.** Temperature dependence of Upara/Uperp =$(U_{33}/(U_{11}/2 + U_{22}/2))$ for the O1, O3 and O4 sites (a) and for the Mo, Fe1 and Fe2 sites in (b).

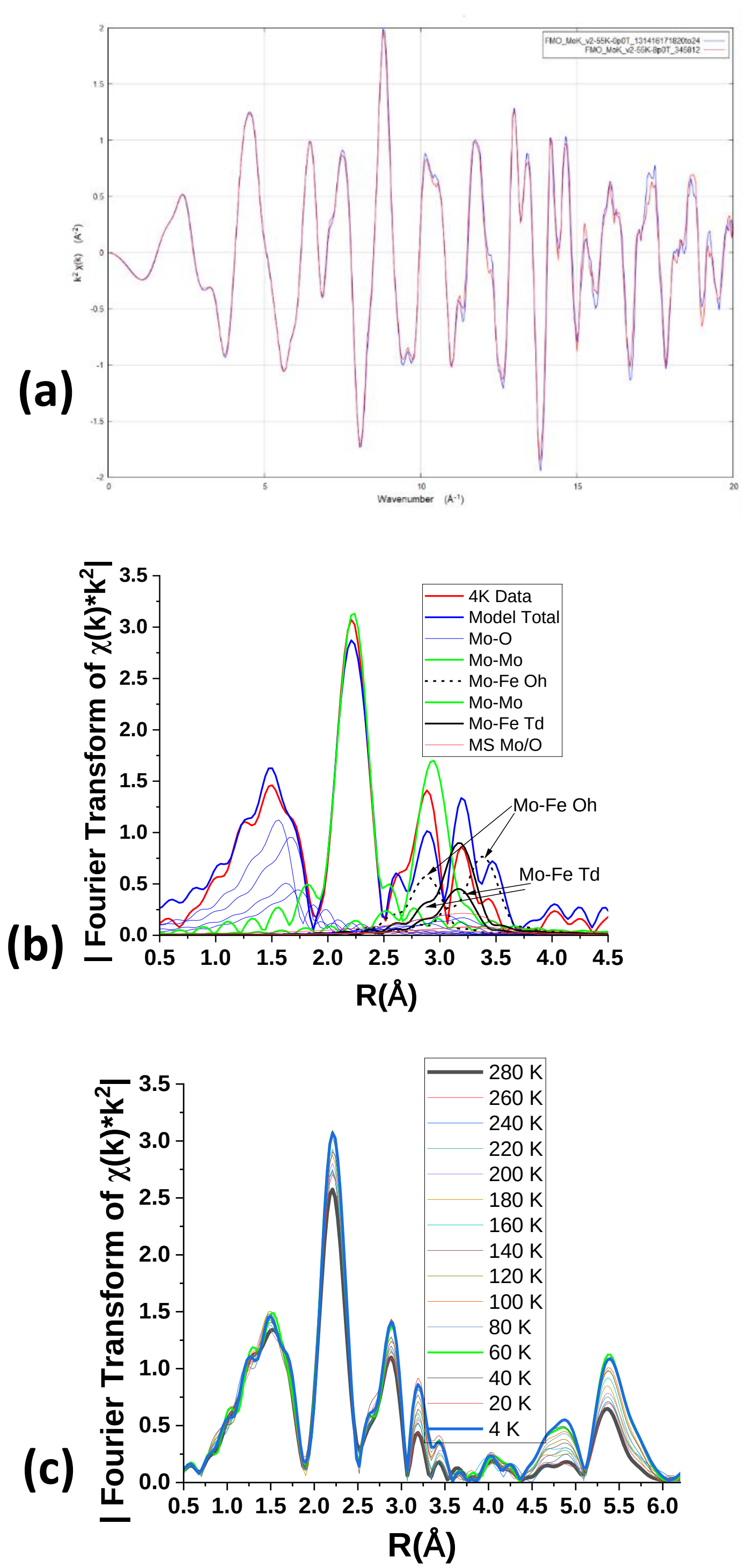

**Fig. S2.** (a) XAFS signal at Mo K-Edge for magnetic fields of 0 T (blue) and 8 T (red) at 55 K. Note the suppression of the amplitude at high k for the 8 T curve. The same is seen in the Fe K-edge XAFS data. (b) Mo K-edge Model XAFS Fourier transform spectrum and its components with MS indicating multiple scattering contributions (weak) based on Feff theoretical calculations [9]. A global Debye-Waller factor of $\sigma^2 = 0.0011$ Å$^2$ was used for all model signals. The red curve is from the measured spectra taken at 4 K. (c) Temperature dependence of Fourier-transformed XAFS spectra between 4 K and 280 K.

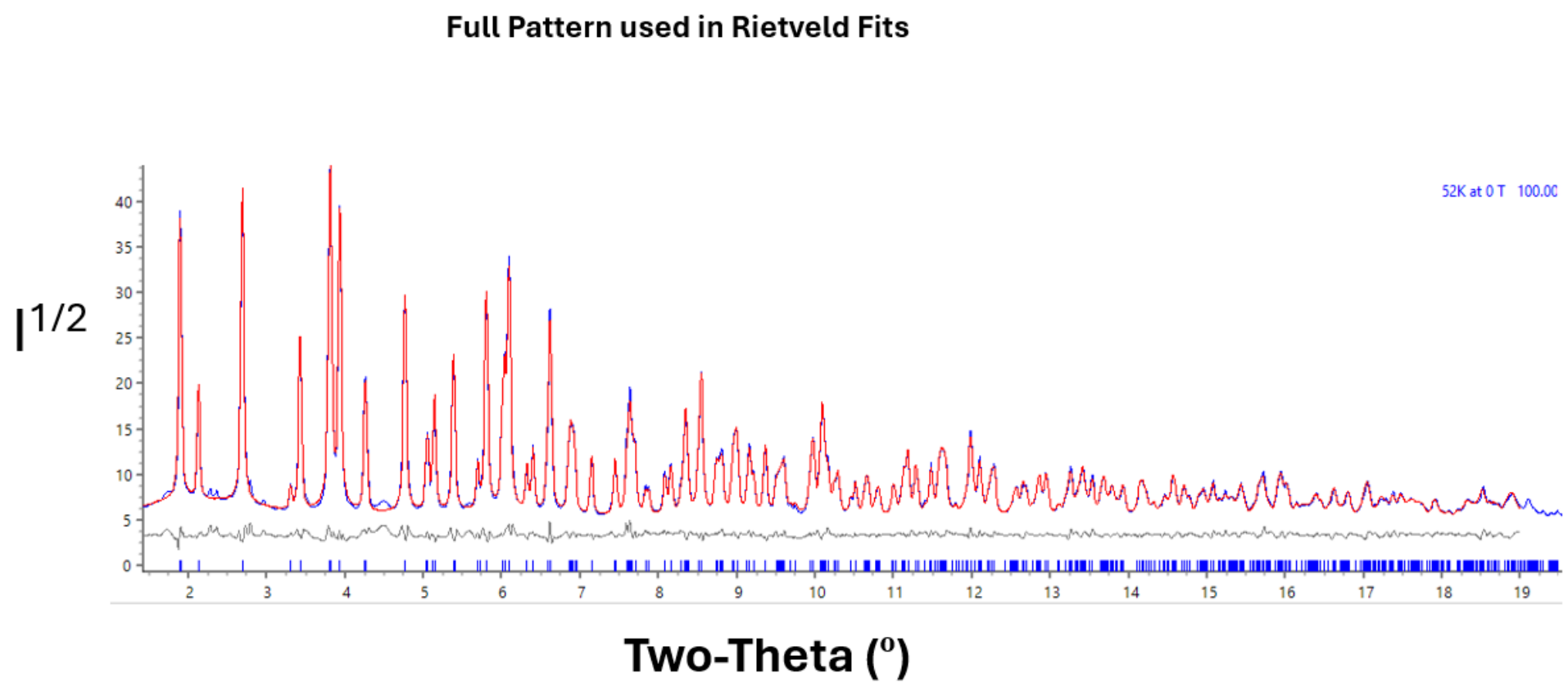

**(a)**

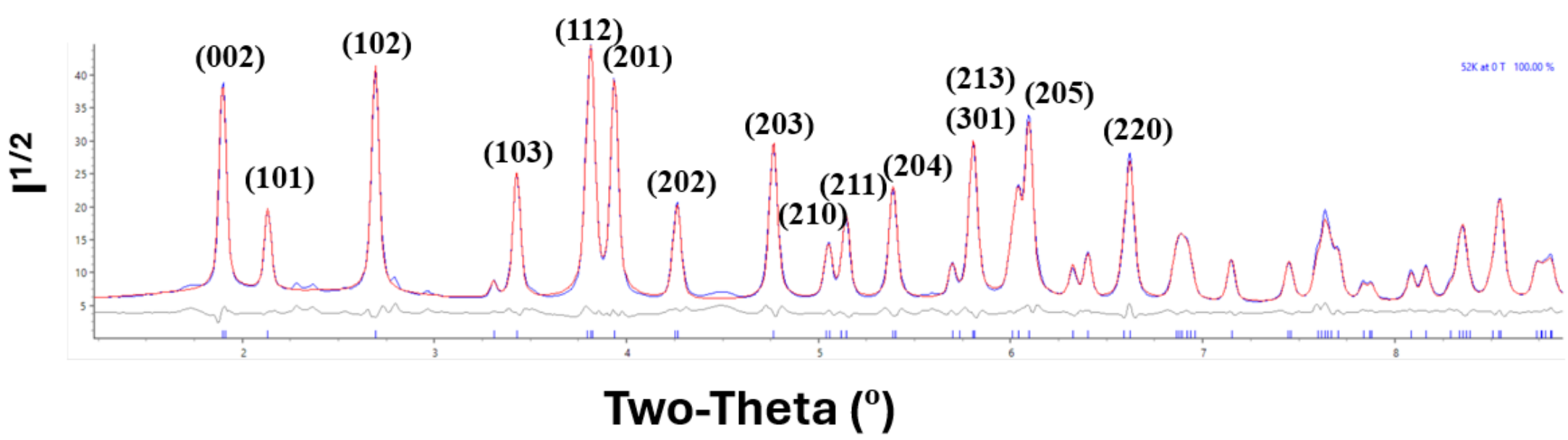

**(b)**

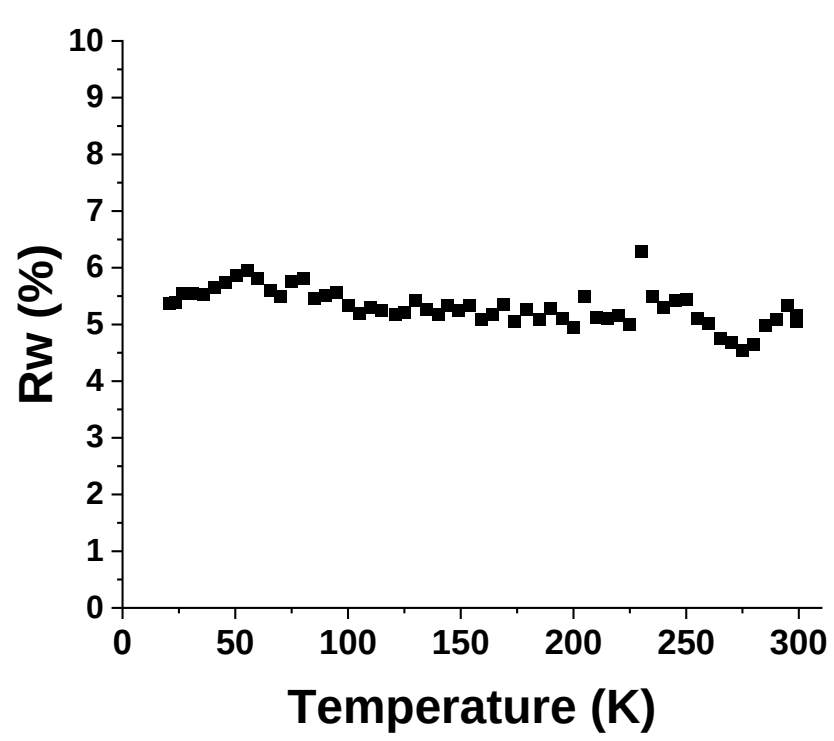

**(c)**

**Fig. S3.** (a) Rietveld fit at a sample temperature of 55 K (0 T, cold finger control temperature = 52 K) over the range $1.0° \leq 2\theta \leq 19°$ and (b) expanded view with peaks labeled by $(hkl)$. Note that the wavelength is 0.1665 Å. The penetration depth (attenuation length) at this wavelength is 1.2 mm. This compares to 10 μm for Cu Kα radiation and 90 μm for Mo Kα radiation used in standard laboratory-based systems. The data are in blue, the red curve is the model fit, and the residual is the lower curve. Blue tick marks on the x-axis are positions of the predicted Bragg peaks based on the model fit. (c) Rietveld fit quality parameter Rw shows no change in space group with temperature over the full range of temperature 4 K to 300 K (warming data). Note that the y-axes for the diffraction patterns in (a) and (b) are plotted as $I^{1/2}$ to accentuate weak features. No new peaks are seen in the fits not conforming to the 300 K space group at ambient pressure and magnetic field.

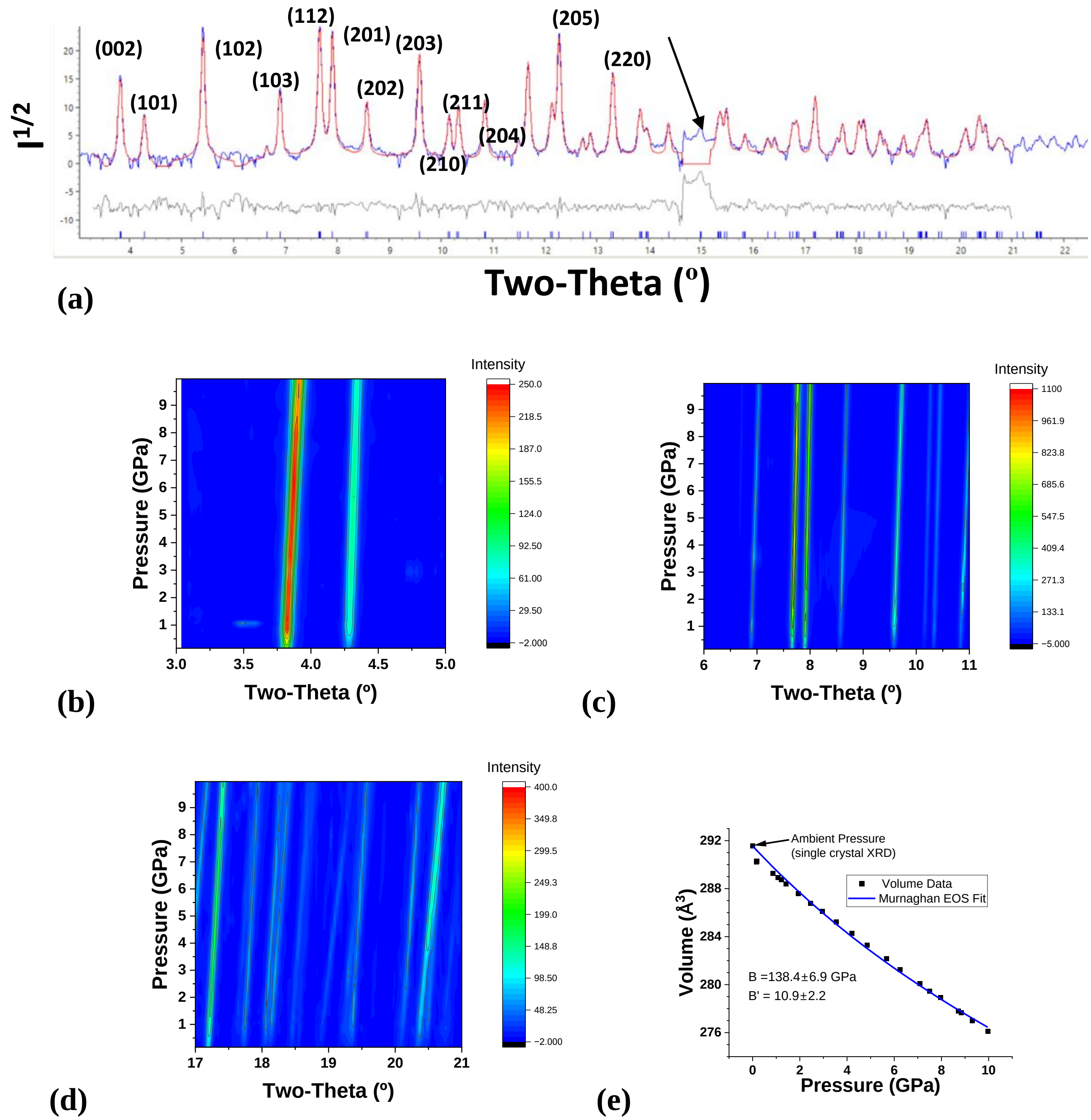

**Fig. S4.** (a) Example ambient temperature high-pressure data set at 1.21 GPa. The data were fit with a LeBail model. The region $14.5° \leq 2\theta \leq 15.2°$ was excluded (arrow). Note that the wavelength is 0.3344 Å for these measurements. The data are in blue, the red curve is the model fit and the residual is the lower curve (grey). Blue tick marks on the x-axis are positions of the predicted Bragg peaks based on the model fit. Note the y-axis for the diffraction pattern is plotted as $I^{1/2}$ to accentuate weak features. Surface plot of data in the low-angle range ((b) and (c)) and the high-angle range (d), revealing no change in structural phase with pressure between ~ambient pressure and 10 GPa. Note that there are no splittings of peaks, discontinuous peak change with pressure, or changes in slope vs pressure in the surface plots. (e) Fit of the volume vs pressure to the Murnaghan equation of state (EOS). B and B' are the extracted bulk modulus and its pressure derivative.

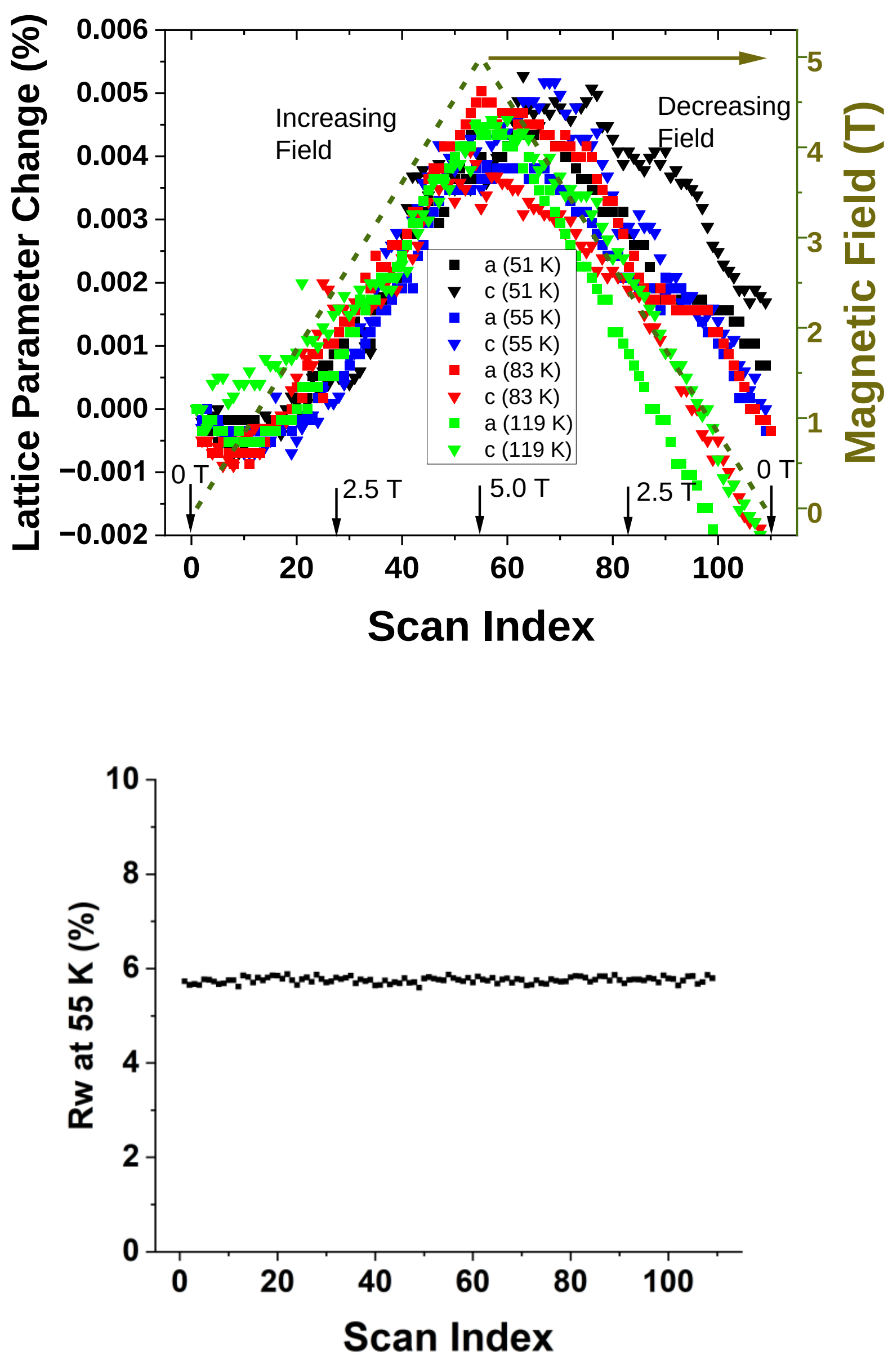

(a)

(b)

**Fig. S5.** Magnetic field-dependent changes in a (squares) and c lattice parameters at 51 K (black), 55 K (blue), 83 K (red), and 119 K (green). The dashed triangle indicates the magnetic field strength (y-label on right). The results are from full Rietveld refinement of powder data as in Fig. S2(a) for a wavelength of 0.1665 Å. For the linear scan index, scan 1 is 0 T while scan 55 is 5T. The field is then ramped down at the same rate after scan 55. Arrows on the x-axis indicate specific magnetic field values. (b) Rietveld fit quality parameter Rw at 55K shows no change in the space group with the magnetic field variations. No new peaks are seen in the fits not conforming to the 300 K space group at ambient pressure and magnetic field.

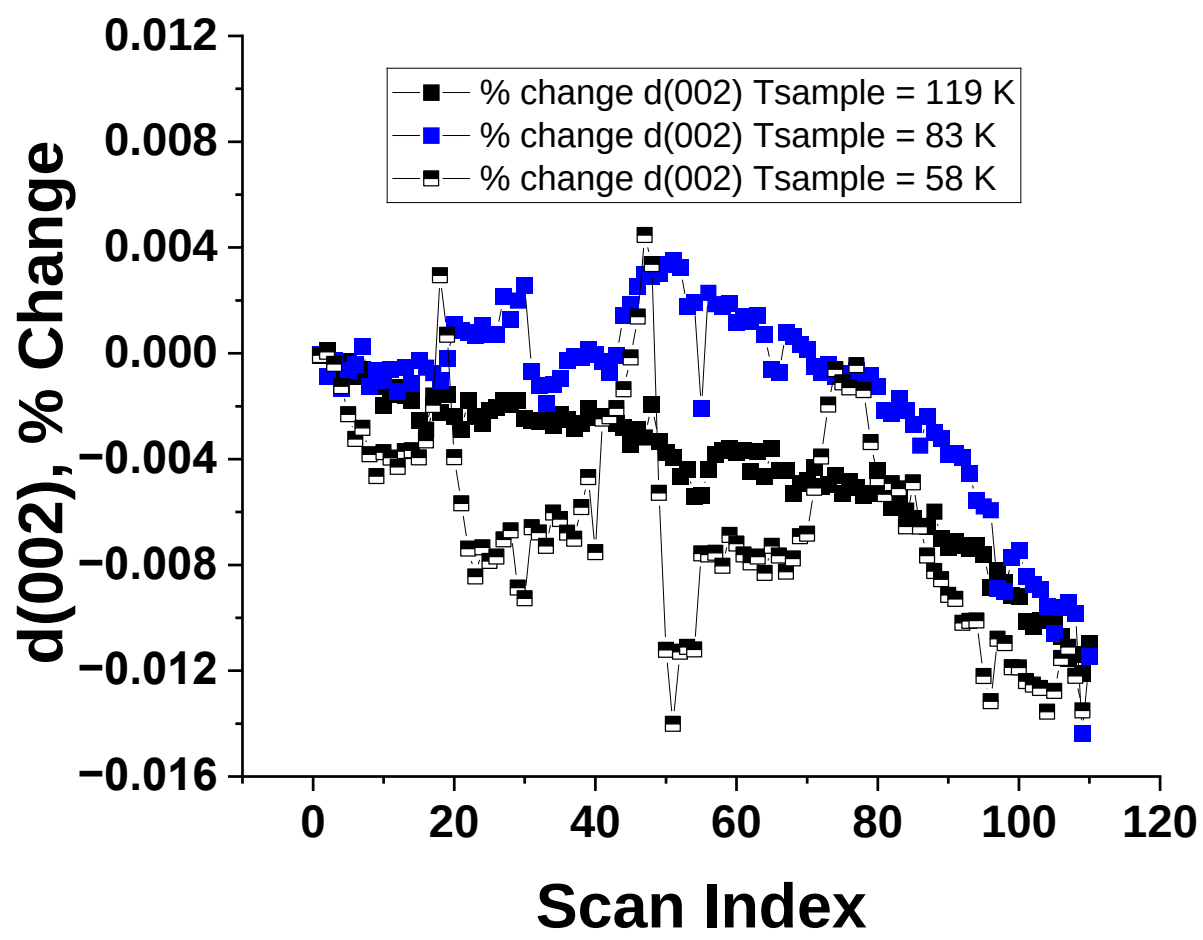

**(a)**

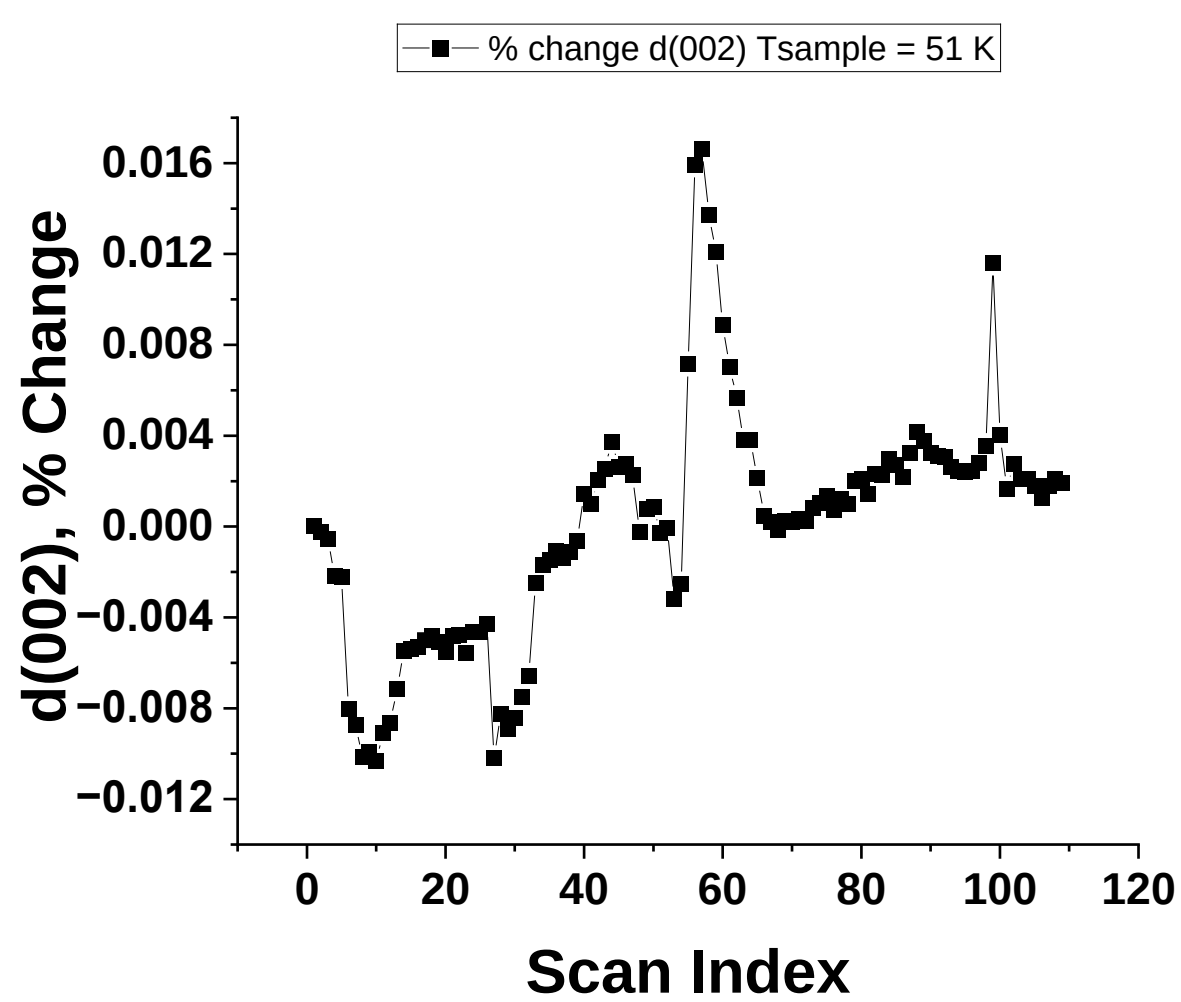

**(b)**

**Fig. S6.** (a) Change in the $d_{002}$ length as a function of magnetic field for increasing field up to 5 T and decreasing field at 58 K, 83 K, and 119 K. The same curve taken at 52 K is given in (b). The high asymmetry with ramping up (scan 0 to 55) relative to ramping down (above scan 55) is due to the first-order nature of the structural change. The x-axis has the same meaning as that in S4(a).